# *Euclid* preparation

## XXX. Performance assessment of the NISP red grism through spectroscopic simulations for the wide and deep surveys

Euclid Collaboration: L. Gabarra[1,2], C. Mancini[3], L. Rodriguez Muñoz[1], G. Rodighiero[1,4], C. Sirignano[1,2],
M. Scodeggio[3], M. Talia[5,6], S. Dusini[2], W. Gillard[7], B. R. Granett[8], E. Maiorano[6], M. Moresco[5,6], L. Paganin[9,10],
E. Palazzi[6], L. Pozzetti[6], A. Renzi[1,2], E. Rossetti[11], D. Vergani[6], V. Allevato[12], L. Bisigello[1,4], G. Castignani[5,6],
B. De Caro[2,1], M. Fumana[3], K. Ganga[13], B. Garilli[3], M. Hirschmann[14,15], F. La Franca[16], C. Laigle[17],
F. Passalacqua[1,2], M. Schirmer[18], L. Stanco[2], A. Troja[1,2], L. Y. A. Yung[19], G. Zamorani[6], J. Zoubian[7], S. Anselmi[1,2],
F. Oppizzi[1,2], G. Verza[1,2], N. Aghanim[20], A. Amara[21], N. Auricchio[6], M. Baldi[5,6,22], R. Bender[23,24], C. Bodendorf[23],
D. Bonino[25], E. Branchini[9,26], M. Brescia[27], J. Brinchmann[28], S. Camera[29,30,25], V. Capobianco[25], C. Carbone[3],
J. Carretero[31,32], F. J. Castander[33,34], M. Castellano[35], S. Cavuoti[12,36], R. Cledassou[37,38], G. Congedo[39],
C. J. Conselice[40], L. Conversi[41,42], Y. Copin[43], L. Corcione[25], A. Costille[44], F. Courbin[45], A. Da Silva[46,47],
H. Degaudenzi[48], J. Dinis[47,46], F. Dubath[48], X. Dupac[41], A. Ealet[43], S. Farrens[49], S. Ferriol[43], M. Frailis[15],
E. Franceschi[6], P. Franzetti[3], S. Galeotta[15], B. Gillis[39], C. Giocoli[6,50], A. Grazian[4], F. Grupp[23,24], L. Guzzo[51,8,52],
W. Holmes[53], A. Hornstrup[54,55], P. Hudelot[56], K. Jahnke[18], M. Kümmel[24], S. Kermiche[7], A. Kiessling[53],
M. Kilbinger[49], T. Kitching[57], R. Kohley[41], B. Kubik[43], M. Kunz[58], H. Kurki-Suonio[59,60], S. Ligori[25], P. B. Lilje[61],
I. Lloro[62], O. Mansutti[15], O. Marggraf[63], K. Markovic[53], F. Marulli[5,6,22], R. Massey[64], S. Maurogordato[65], S. Mei[13],
M. Meneghetti[6,22], G. Meylan[45], L. Moscardini[5,6,22], E. Munari[15], R. C. Nichol[66], S.-M. Niemi[67], J. Nightingale[64],
C. Padilla[31], S. Paltani[48], F. Pasian[15], K. Pedersen[68], W. J. Percival[69,70,71], V. Pettorino[49], G. Polenta[72], M. Poncet[37],
F. Raison[23], J. Rhodes[53], G. Riccio[12], E. Romelli[15], M. Roncarelli[6], R. Saglia[24,23], D. Sapone[73], P. Schneider[63],
A. Secroun[7], G. Seidel[18], S. Serrano[34,74], G. Sirri[22], C. Surace[44], P. Tallada-Crespí[75,32], D. Tavagnacco[15],
A. N. Taylor[39], I. Tereno[46,76], R. Toledo-Moreo[77], F. Torradeflot[75,32], I. Tutusaus[78], E. A. Valentijn[79],
L. Valenziano[6,22], T. Vassallo[15], Y. Wang[80], J. Weller[24,23], A. Zacchei[15,81], S. Andreon[8], H. Aussel[82], S. Bardelli[6],
M. Bolzonella[6], A. Boucaud[13], E. Bozzo[48], C. Colodro-Conde[83], D. Di Ferdinando[22], M. Farina[84], J. Graciá-Carpio[23],
E. Keihänen[85], V. Lindholm[59,60], D. Maino[51,3,52], N. Mauri[86,22], Y. Mellier[17,87], C. Neissner[31], V. Scottez[56,88],
M. Tenti[89], E. Zucca[6], Y. Akrami[90,91,92,93,94], C. Baccigalupi[95,81,15,96], M. Ballardini[97,98,6], F. Bernardeau[99,17],
A. Biviano[15,81], A. S. Borlaff[100], E. Borsato[1,2], C. Burigana[97,101,89], R. Cabanac[78], A. Cappi[6,65], C. S. Carvalho[76],
S. Casas[102], T. Castro[81,15,96], K. Chambers[103], A. R. Cooray[104], J. Coupon[48], H. M. Courtois[105], S. Davini[10],
S. de la Torre[44], G. De Lucia[15], G. Desprez[48,106], H. Dole[20], J. A. Escartin[23], S. Escoffier[7], I. Ferrero[61], F. Finelli[6,89],
S. Fotopoulou[107], J. Garcia-Bellido[90], K. George[108], G. Giacomini[22], G. Gozaliasl[59], H. Hildebrandt[109], I. Hook[110],
O. Ilbert[44], A. Jimenez Muñoz[111], J. J. E. Kajava[112], V. Kansal[82], C. C. Kirkpatrick[85], L. Legrand[58], A. Loureiro[39,94],
J. Macias-Perez[111], M. Magliocchetti[84], G. Mainetti[113], R. Maoli[114,35], S. Marcin[115], M. Martinelli[35,116], N. Martinet[44],
C. J. A. P. Martins[117,28], S. Matthew[39], L. Maurin[20], R. B. Metcalf[5,6], G. Morgante[6], S. Nadathur[21],
A. A. Nucita[118,119,120], L. Patrizii[22], V. Popa[121], C. Porciani[63], D. Potter[122], M. Pöntinen[59], A. G. Sánchez[23],
Z. Sakr[78,123,124], A. Schneider[122], E. Sefusatti[15,96,81], M. Sereno[6,22], A. Shulevski[125,79], A. Spurio Mancini[57],
J. Stadel[122], J. Steinwagner[23], R. Teyssier[126], J. Valiviita[59,60], A. Veropalumbo[51],
M. Viel[95,81,15,96], and I. A. Zinchenko[24]

*(Affiliations can be found after the references)*

## ABSTRACT

This work focusses on the pilot run of a simulation campaign aimed at investigating the spectroscopic capabilities of the *Euclid* Near-Infrared Spectrometer and Photometer (NISP), in terms of continuum and emission line detection in the context of galaxy evolutionary studies. To this purpose, we constructed, emulated, and analysed the spectra of 4992 star-forming galaxies at $0.3 \leq z \leq 2.5$ using the NISP pixel-level simulator. We built the spectral library starting from public multi-wavelength galaxy catalogues, with value-added information on spectral energy distribution (SED) fitting results, and stellar population templates from Bruzual & Charlot (2003, MNRAS, 344, 1000). Rest-frame optical and near-IR nebular emission lines were included using empirical and theoretical relations. Dust attenuation was treated using the Calzetti extinction law accounting for the differential attenuation in line-emitting regions with respect to the stellar continuum. The NISP simulator was configured including instrumental and astrophysical sources of noise such as the dark current, read-out noise, zodiacal background, and out-of-field stray light. In this preliminary study, we avoided contamination due to the overlap of the slitless spectra. For this purpose, we located the galaxies on a grid and simulated only the first order spectra. We inferred the $3.5\sigma$ NISP red grism spectroscopic detection limit of the continuum measured in the $H$ band for star-forming galaxies with a median disk half-light radius of $0\farcs4$ at magnitude $H = 19.5 \pm 0.2$ AB mag for the *Euclid* Wide Survey and at $H = 20.8 \pm 0.6$ AB mag for the *Euclid* Deep Survey. We found a very good agreement with the red grism emission line detection limit requirement for the Wide and Deep surveys. We characterised the effect of the galaxy shape on the detection capability of the red grism and highlighted the degradation of the quality of the extracted spectra as the disk size increased. In particular, we found that the extracted emission line signal-to-noise ratio (S/N) drops by ~45% when the disk size ranges from $0\farcs25$ to $1''$. These trends lead to a correlation between the emission line S/N and the stellar mass of the galaxy and we demonstrate the effect in a stacking analysis unveiling emission lines otherwise too faint to detect.

## 1. Introduction

The *Euclid* mission (Laureijs et al. 2011; Racca et al. 2016) aims to study the dark Universe with two primary cosmological probes, weak lensing (WL) and galaxy clustering (GC). To meet its primary science goals, the 6-yr *Euclid* observation scheme includes ~14 500 deg$^2$ of the extra-galactic sky using both multi-band imaging and low-resolution slitless spectroscopy. The resulting survey is the *Euclid* Wide Survey (EWS; Scaramella et al. 2022) and will be complemented with the *Euclid* Deep Survey (EDS; Scaramella et al., in prep), which will cover ~ 40 deg$^2$ and reach two magnitudes fainter.

*Euclid* has been designed to probe the Universe since redshift ~2, which corresponds to the cosmic time when dark energy started to drive the accelerating expansion of the Universe (Amendola & Tsujikawa 2010). The survey will be particularly sensitive to the redshift range $0.9 \leq z \leq 1.8$, where the H$\alpha$ line falls in the EWS spectroscopic channel passband with several thousands of sources per square degree (see Pozzetti et al. 2016, Fig. 4) within the *Euclid* spectroscopic sensitivity (Scaramella et al. 2022). This period includes 'cosmic noon' (Madau et al. 1998; Madau & Dickinson 2014) when galaxies were particularly prolific in terms of the star-formation rate (SFR), which has been declining ever since. It also covers the redshift range $1.4 \leq z \leq 1.8$ known as the 'redshift desert'. This redshift range has poor coverage in current spectroscopic surveys due to the strong atmospheric absorption in the near-IR where the strongest galactic optical emission and absorption spectral features are redshifted, while the strong UV features are still too blue to be observed with optical spectrographs (see Le Fèvre et al. 2015, Fig. 13).

To fulfil its objectives, two instruments sharing a field of view of ~0.5 deg$^2$ will be mounted on *Euclid*. First, the visual imager (VIS; Cropper et al. 2016) with a spatial resolution of $0\farcs18$ and outfitted with one single filter covering the $I_E$ band (0.55–0.90 µm) will serve to measure cosmic shear for the WL probe. Second, the Near-Infrared Spectrometer and Photometer (NISP; Maciaszek et al. 2016) will carry out photometry (NISP-P) and spectrometry (NISP-S). The NISP-P channel includes three broadband filters in the filter wheel assembly, $Y_E$ (0.95–1.21 µm), $J_E$ (1.17–1.57 µm), and $H_E$ (1.52–2.02 µm), with a spatial resolution of $0\farcs3$ in all three bands (Schirmer et al. 2022). The NISP-S includes two grisms in a grism wheel assembly, the red grism (RGS) covering the $RG_E$ band (1.25–1.86 µm) and the blue grism (BGS), which will only be used in the EDS covering the $BG_E$ band (0.92–1.30 µm). The NISP-S channel is designed to provide an accurate redshift determination for emission line galaxies with $\sigma_z/(1 + z) \leq 0.001$.

Beyond the excellent performance expected from the cosmological probes, the unprecedented volume of spectro-photometric data including accurate morphological parameters for billions of galaxies and tens of millions of spectroscopic redshifts will be of great interest for legacy science purposes. In particular, the *Euclid* dataset will enable the community to study scaling relations, mass assembly, environmental effects, and galaxy-active galactic nucleus (AGN) co-evolution on samples including a large number of massive star-forming and passive galaxies at intermediate and high redshift.

In this paper, we focus our study on star-forming galaxies (SFGs) using the pixel simulator of the RGS channel that has been parameterised with the NISP optical performance evaluated during the ground-test campaigns (Waczynski et al. 2016; Barbier et al. 2018; Costille et al. 2019; Maciaszek et al. 2022). The simulator includes instrumental noise, for example a dark current and read-out noise, as well as astrophysical noise, for example zodiacal background and out-of-field stray light. This effort, which is part of the *Euclid* legacy science programme, is referred as the pilot run and is the first step of a simulation campaign. The pilot run includes simulations of thousands of SFG first order spectra, which were positioned on a grid to avoid contamination between spectra. It will be followed in the near future by the full run, which will simulate tens of thousands of spectra of a wide variety of galaxy types, for example AGNs, passive galaxies, and SFGs, and including the assessment of the contamination due to the overlapping spectra. The main objectives of our efforts have been the following: i) to develop a solid methodology to build a realistic and representative synthetic spectral energy distribution (SED) library of SFGs at $0.3 \leq z \leq 2.5$[1] (hereafter referred to as the incident spectra), including reliable emission line fluxes and widths[2]; ii) to test the effect of the galaxy shape on slitless spectroscopy; and iii) to provide a preliminary assessment on the continuum and emission line detection capabilities of the RGS channel.

We present in Sect. 2 the data and sample selection for which SED-fitting parameters are available in publicly released catalogues. In Sect. 3, we provide a detailed procedure for constructing the incident spectra. This process begins with a template continuum, to which we incorporate the flux predictions of the nebular emission lines. The frame of the pilot run and setup of the *Euclid* NISP-S pixel-level simulator, that is TIPS (Zoubian et al. 2014), is presented in Sect. 4. The analysis of the 1D extracted spectra is presented in Sect. 5. A summary of the caveats of the simulations is presented in Sect. 6. The conclusion and main results are presented in Sect. 7.

---
[1] This redshift range has been chosen to probe multiple strong emission lines, such as the [S III]$\lambda$9531 at $z \geq 0.3$ and [O III]$\lambda$5008 up to $z = 2.5$.
[2] The spectral libraries are available upon request.

We adopted a $\Lambda$CDM[3] cosmology with $\Omega_m = 0.3$, $\Omega_\Lambda = 0.7$, and $H_0 = 70 \, \text{km s}^{-1} \, \text{Mpc}^{-1}$. Except if stated otherwise, we adopted in this work the extinction curve from Calzetti et al. (2000) and the parameterisation of the initial mass function (IMF) proposed by Chabrier (2003). All magnitudes are expressed in the AB system (Oke & Gunn 1983).

## 2. Data and sample selection

### 2.1. Preparation of the Euclid Wide Survey simulation with the COSMOS2015 catalogue

For the EWS simulation, we relied on the public multi-wavelength spectro-photometric catalogue COSMOS2015 (Laigle et al. 2016, hereafter L16) that covers the COSMOS field (Scoville et al. 2007) and that we complemented with estimated morphological parameters as described in Sect. 2.5. This catalogue contains an updated version of the photometric redshifts, together with estimates of parameters including the stellar mass, dust extinction, and SFR obtained from SED-fitting for more than half a million sources.

The COSMOS survey includes several spectroscopic surveys including 3D-HST (Brammer et al. 2012; Momcheva et al. 2016), FMOS (Kashino et al. 2019), and Large Early Galaxy Astrophysics Census (LEGA-C; Van Der Wel et al. 2016), which provide measurements of some of the emission lines of interest in this study (see Sect. 3.4.2). The area of 1.7 deg$^2$ covered by the L16 catalogue includes those rare, very massive, and bright galaxies at intermediate and high redshift that will represent the majority of spectroscopic EWS targets. The L16 catalogue was therefore chosen for the EWS simulation.

We selected sources with magnitude $17 \leq H \leq 24$ and at $0.3 \leq z \leq 2.5$. These limits were chosen to reach the EWS NISP-P magnitude limit of $H_E = 24 \, \text{mag}$[4] (Scaramella et al. 2022), and to cover a redshift range that enabled us to probe multiple strong emission lines. Furthermore, since the target of this study are line emitters, we selected the sub-samples to be simulated among SFGs based on colour-colour diagrams available in the literature. The NUV, $r$, $J$ absolute magnitudes are available in the L16 catalogue, we therefore referred to the identification criteria for SFGs proposed by Laigle et al. (2016) using the colour-colour NUV$rJ$ diagram. These selection criteria provided us with 156 323 sources.

### 2.2. Preparation of the Euclid Deep Survey simulation with the BARRO2019 catalogue

For the EDS simulation, we relied on the public multi-wavelength spectro-photometric catalogue BARRO2019 (Barro et al. 2019, hereafter B19) released by the Cosmic Assembly Near-infrared Deep Extragalactic Legacy Survey (CANDELS; Grogin et al. 2011; Koekemoer et al. 2011) that covers the Great Observatories Origins Deep Survey-North field (GOODS-N; Giavalisco et al. 2004). We used the SED-fitting parameters available in the B19 catalogue, that we complemented with results of Sérsic profile fits from CANDELS imaging mosaics (Van Der Wel et al. 2012), and the bulge-disk decomposition from Dimauro et al. (2018). The B19 catalogue includes spectroscopic measurements for some of the sources using the HST/WFC3 grisms G102/G141.

The area of ∼160 arcmin$^2$ covered by the B19 catalogue consists of imaging data of galaxies at intermediate stellar mass including, on average, fainter galaxies compared to the L16 catalogue. The B19 catalogue was therefore chosen for the EDS simulation.

We selected sources with $17 \leq H \leq 26$ and at $0.3 \leq z \leq 2.5$ as similarly done for the L16 catalogue but going two magnitudes deeper to reach the EDS NISP-P magnitude limit at $H_E = 26 \, \text{mag}$ (Racca et al. 2016).

To select SFGs, we proceeded using colour-colour diagrams in a similar way as for the EWS simulation but using the $U$, $V$, $J$ absolute magnitudes available in the B19 catalogue. We referred to the identification criteria proposed by Williams et al. (2009) using the colour-colour $UVJ$ diagram. These selection criteria provided us with 15 460 sources.

### 2.3. Selection of galaxies to be simulated

To avoid the spectra being overlapped, we set an upper limit of 2496 galaxies per pointing that we located on the 16 detectors of the NISP focal plane (see Sect. 4 for further details on the pixel-level simulator configuration). The emission lines flux predictions used to select a representative sample of galaxies targetted by NISP are presented in Sect. 3.2.

#### 2.3.1. Sample selection to be simulated for the EWS and EDS simulations

To create a sample representative of the diversity of objects available in the catalogues, we made use of a three dimensional grid of redshift, total stellar mass, and flux of the brightest emission line falling in the $RG_E$ band.

Referring to the EWS requirement in terms of emission line detection with a signal-to-noise ratio (S/N) equal to 3.5 for the H$\alpha$ line of a 0$''$25 radius source located at $z = 1.4$, that is $\lambda = 16\,000$ Å, set at $2 \times 10^{-16} \, \text{erg s}^{-1} \, \text{cm}^{-2}$ (Scaramella et al. 2022), and at $6 \times 10^{-17} \, \text{erg s}^{-1} \, \text{cm}^{-2}$ for the EDS, we set the lower limit for the sample to be simulated at $1 \times 10^{-16} \, \text{erg s}^{-1} \, \text{cm}^{-2}$ for the EWS simulation and at $4.5 \times 10^{-17} \, \text{erg s}^{-1} \, \text{cm}^{-2}$ for the EDS simulation to be able to probe the detection sensitivity at its limits. We set the limit at two times below the requirement for the EWS simulation to offer candidate sources to study the potential of stacking analyses. We used as references the three strongest lines expected to be observed at different redshifts: [S III]$\lambda$9531 at $0.3 \leq z \leq 0.9$, H$\alpha$ at $0.9 \leq z \leq 1.82$ and [O III]$\lambda$5008 at $1.5 \leq z \leq 2.5$ (see Fig. 1).

As presented in Sect. 2, we used sources from the L16 catalogue for the EWS simulation (red triangles in Fig. 1). For the EDS simulation, we used sources from the B19 catalogue (blue triangles) that we complemented with sources from the L16 (orange triangles) to be able to reach the capacity of 2496 sources covering a wider range of physical parameters while including more massive and stronger emitter galaxies. We can see in the middle panel of Fig. 1 that at low mass, some sources with an H$\alpha$ flux below the sample limit are kept. A similar effect for massive galaxies is visible on the right panel where we kept some sources that have an [O III]$\lambda$5008 flux below the sample limit. This is due to the fact that at redshift $1.5 \leq z \leq 1.82$, both H$\alpha$ and [O III]$\lambda$5008 fall in the $RG_E$ band. Therefore, for a source located in this redshift range, if one of the two emission lines has a flux above the sample limit then the source has been kept even if the flux of the other emission line is below the sample limit.

---

[3] $\Lambda$CDM is a widely accepted theoretical framework in modern cosmology. The term $\Lambda$ refers to the cosmological constant that represents the energy density of dark energy, while "CDM" refers to cold dark matter.

[4] This choice to use the photometric limit for the spectroscopic simulations was motivated by the fact that a spectrum will be extracted for all detected photometric sources. The same applies for the *Euclid* Deep Survey catalogue presented in Sect. 2.2.

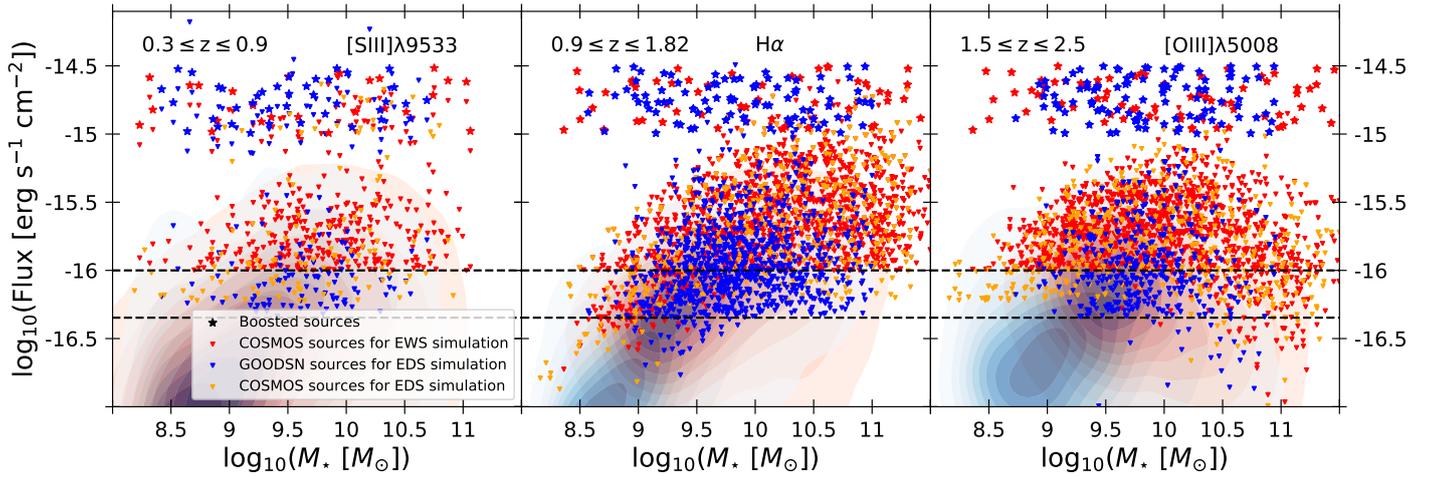

**Fig. 1.** Distribution of emission line fluxes as a function of the mass. This figure presents the selected sources to be simulated considering the limits set for the EWS and EDS in terms of the emission line detection limits. The sources represented by stars are the sources with boosted emission line flux (explained in the text). The limits are set below the respective *Euclid* requirements in order to characterise emission line detection capability around these requirements. The limits have been set at $1 \times 10^{-16}$ erg s$^{-1}$ cm$^{-2}$ (top dashed black line) for the EWS simulation and at $4.5 \times 10^{-17}$ erg s$^{-1}$ cm$^{-2}$ (bottom dashed black line) for the EDS simulation. The red and blue shaded regions correspond to iso-proportions of the distribution density of the L16 and B19 catalogues, respectively, starting at 20% with a 10% step. The red triangles are sources from L16 selected for the EWS simulation. The blue triangles are sources from B19 selected for the EDS simulation, which has been completed with sources from L16 (orange triangles) to reach the maximum number of sources possible in a simulated pointing. *Left*: distribution of the [S III]$\lambda$9533 fluxes in the redshift range allowing detection with the RGS. *Middle*: distribution of the H$\alpha$ fluxes. *Right*: distribution of the [O III]$\lambda$5008 fluxes.

For calibration purposes, we boosted the continuum by $-3.7$ mag and the emission line fluxes to reach a value normally distributed between $10^{-15}$ and $3 \times 10^{-15}$ erg s$^{-1}$ cm$^{-2}$ for about 5% of the sources of the sample. These sources are indicated with star markers on Fig. 1.

### 2.3.2. Sample to be simulated for the study of the morphological effects on the slitless spectroscopy performance

While the spectral resolution of long-slit spectroscopy depends in part on the slit width, the resolution of slitless spectroscopy depends on the object shape and more specifically on the object size along the dispersion axis (Pasquali et al. 2006; Kümmel et al. 2009). It is therefore crucial to understand the effect of the object shape on the quality of the NISP spectra. However, the disk R50 distribution of the samples selected to simulate the EWS and EDS peak at disk R50 = 0″.3, and the different variables that can affect the spectra makes it difficult to isolate the effect from a single morphological parameter. For this purpose, we have explored the effects related to different morphological parameters given as input to TIPS (see Sect. 2.5) by constructing a dedicated sample. To probe the extent to which the galaxy morphology affects the extracted spectra, TIPS was configured in the following manner. We chose a bright source located at $z = 1.6$, where the two strongest emission lines, H$\alpha$ and [O III]$\lambda$5008, fall in the $RG_E$ band. The respective fluxes of the lines are H$\alpha = 6 \times 10^{-16}$ erg s$^{-1}$ cm$^{-2}$ and [O III]$\lambda$5008 = $3.5 \times 10^{-16}$ erg s$^{-1}$ cm$^{-2}$ with magnitude $H = 19$ mag. The morphological parameters were changed one at a time, while the others were kept at their default value. The default values of the morphological parameters to be tested are: disk R50 = 0″.5, bulge fraction = 0.2, inclination angle = 45°, position angle = 45°. We then obtained four sub-samples of 1248 sources in which only one parameter varies, enabling us to disentangle the effect of these parameters on the quality of the extracted spectra. The range of variation are 0.1–2″ for the disk R50, 0–1 for the bulge fraction, and 0–90° for the inclination and position angles.

### 2.4. The $M_\star$-SFR plane and the main sequence

It is interesting to characterise our sample in terms of star-formation rate (SFR). It is well known that SFGs lie on a tight sequence in the total stellar mass ($M_\star$) and SFR plane, referred as the 'main sequence' (MS; Noeske et al. 2007; Daddi et al. 2007; Wuyts et al. 2011; Rodighiero et al. 2011, 2014; Popesso et al. 2022, and many others). It has been shown that the normalisation of the MS is shifted towards higher SFR with increasing redshift reaching a peak at $1.5 \leq z \leq 3$ (Madau et al. 1998; Madau & Dickinson 2014). We show in Fig. 2 the normalisation of the MS derived by Rodighiero et al. (2011) for SFGs at $z \sim 2$, and formulated as follows:

$$\log_{10}(\text{SFR}[M_\odot \, \text{yr}^{-1}]) = -6.42 + 0.79 \log_{10}(M_\star[M_\odot]). \quad (1)$$

We then made use of the redshift dependence of the specific star-formation rate (sSFR), defined as the ratio SFR[$M_\odot$ yr$^{-1}$]/$M_\star$[$M_\odot$], scaling as,

$$\text{sSFR}(z) = \text{sSFR}(0)(1 + z)^{2.8}, \quad (2)$$

where sSFR(0) is the sSFR at $z = 0$ (Sargent et al. 2014). We obtained a redshift-dependent relation that we used up to $z = 2.5$ as follows:

$$\text{SFR}(z_{\text{bin}}) = \text{SFR}(z = 2)[(1 + z_{\text{bin}})/3]^{2.8}. \quad (3)$$

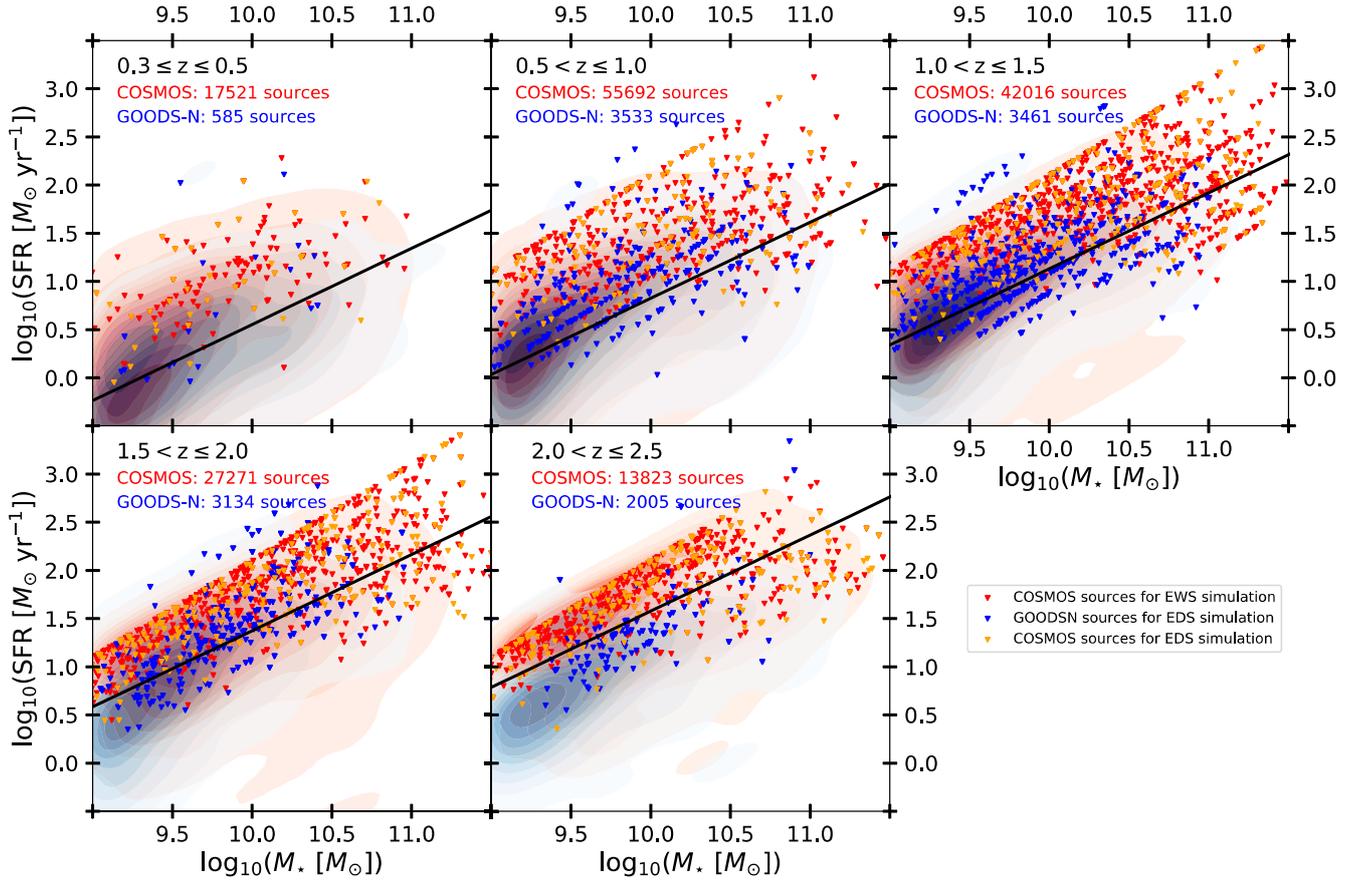

**Fig. 2.** SFR versus total stellar mass diagram for five different redshift bins. The solid black line shows the normalisation of the 'main sequence' from Rodighiero et al. (2011) derived at $z \sim 2$ that we extrapolated to different redshift using the redshift dependence proposed by Sargent et al. (2014). Readers can refer to the caption of Fig. 1 for a colour assignment description.

We indicate in Fig. 2 the sources from the L16 (red shaded region) and from the B19 catalogue (blue shaded region). We also present the final selection of sources to be simulated with triangles. For a description of the final selection process, we refer the reader to Sect. 2.3. The SFR and stellar mass indicated in Fig. 2 are the SED-fitting parameters available in the catalogues. It has been shown by Laigle et al. (2019) that using SED-fitting parameters is robust for the stellar mass but has to be considered with caution for the SFR. It is worth noting that the photometry available in the B19 and L16 differs in terms of covered wavelength range. In particular far infrared data from *Herschel* is available only in the B19 catalogue. However, we decided to use the SFR derived from SED-fitting for consistency with the other parameters. This choice was justified by the good agreement found with the SFR derived from dust-corrected UV photometric measurements for the sources of the B19 catalogue. We can see in Fig. 2 that the SFR of the selected sources for the EDS simulation (blue triangles) are typically below the SFR of the selected sources for the EWS simulation. This comes from our selection procedure, which considers the detection capability of the *Euclid* spectroscopic channel. The EDS will be able to probe galaxies on both side of the main sequence, while the EWS will essentially probe galaxies above the main sequence.

### 2.5. Morphological parameters

Morphological parameters are required by TIPS to produce a simulated image with a realistic surface brightness distribution (see Sect. 4 for a description of the simulator). In particular, we are interested in the following morphological parameters: i) the bulge-to-total mass fraction, ii) the bulge half-light radius (bulge R50), iii) the disk half-light radius (disk R50)[5], iv) the minor to major axis ratio of the bulge ($b/a$), v) the inclination angle of the galaxy (0° = face-on, 90° = edge-on), and vi) the position angle of the galaxy on the sky with respect to the north. As anticipated above, these parameters have been made available for the B19 catalogue by Van Der Wel et al. (2012) using Sérsic profile fits from the CANDELS imaging mosaics and by Dimauro et al. (2018) who performed the bulge-disk decomposition.

To estimate the bulge fraction for the L16 catalogue sources, we referred to the calibration from the stellar mass of the Empirical Galaxy Generator (EGG; Schreiber et al. 2017, Eq. (3)). We derived the bulge R50 from an empirical fit on the parameters from Dimauro et al. (2018) using our estimated bulge fraction. For the disk R50, we used the $M_\star$-size calibration for SFGs proposed by Van Der Wel et al. (2014, see Fig. 3). We define the ratio $b/a$ from the relation $b/a = [\cos^2(\theta_{GAL}) + C^2 \sin^2(\theta_{GAL})]^{0.5}$ with $C = 0.6$ as derived by Rodríguez & Padilla (2013) for local elliptical galaxies. We used this estimation by approximating the bulge of the SFGs with an elliptical galaxy, as done for the construction of the *Euclid* true Universe catalogue. The position and inclination ($\theta_{GAL}$) angles have been set randomly with values ranging from 0° to 90°.

---

[5] We recall here that the relation between the scale length, $R_h$, and the half-light radius, R50, for an exponential disk is $R_h =$ R50/1.678.

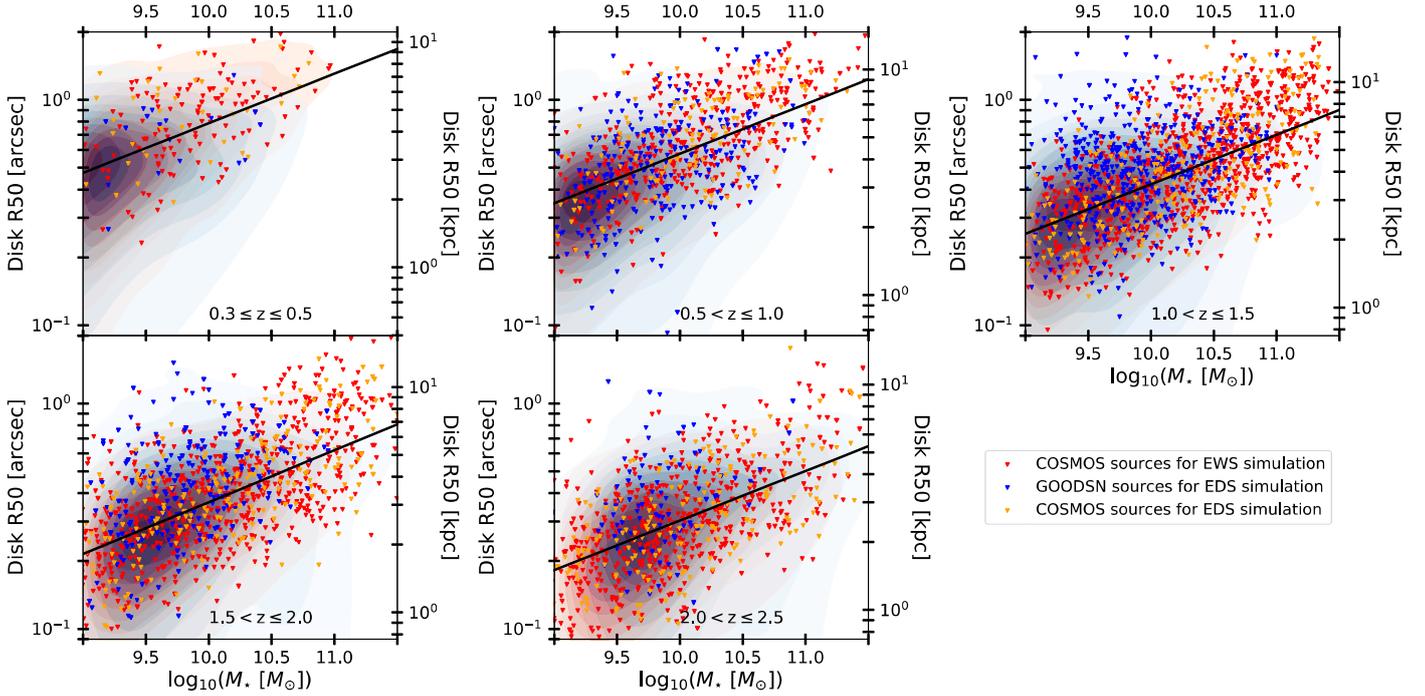

**Fig. 3.** $M_\star$-size relation for five different redshift bins where the size is the disk half-light radius (disk R50) indicated in arcseconds on the left vertical axis and in kiloparsecs on the right vertical axis. The solid black lines show the redshift-dependent normalisation of the $M_\star$-size relation proposed by Van Der Wel et al. (2014). Readers can refer to the caption of Fig. 1 for a colour assignment description.

## 3. Construction of the incident spectra: An observational approach

### 3.1. Construction of the spectral continuum

We made use of available observed data to generate a synthetic continuum using the synthesis code GALAXEV by Bruzual & Charlot (2003). This code allows us to compute the spectral evolution of stellar populations over wide ranges of age and metallicities and to derive synthetic spectra with both low and high resolution (denoted by lr and hr) over the wavelength range from 91 Å to 160 µm. Among those available, we used the high-resolution templates, that is 3 Å, from the STELIB spectroscopic stellar library (LeBorgne et al. 2003), consisting of a homogeneous library of 249 stellar spectra at optical wavelengths (0.32–0.95 µm), and are complemented with lower resolution models at longer wavelengths. We made use of the following physical parameters: the magnitude in the $H$ band, $M_\star$, $V$ band dust attenuation ($A_V$), age, and star-formation history (SFH). These parameters are available for the galaxies in the two catalogues mentioned previously, that is L16 and B19. From both catalogues, we took the SED-fitting parameters inferred using the LePhare software (Arnouts et al. 2002; Arnouts & Ilbert 2011). We assumed Solar metallicity (m62), and we followed the parameterisation by Chabrier (2003) for the IMF. The IMF was chosen for consistency with the IMF that was used to derive the SED fitting results in the adopted catalogues, which is expressed as

$$\phi(m) \propto \begin{cases} \exp\left[-\frac{\log_{10}^2(m/m_c)}{2\sigma^2}\right] & \text{if } m \leq 1\,M_\odot \\ m^{-1.3} & \text{if } m > 1\,M_\odot, \end{cases} \quad (4)$$

where $m_c = 0.08\,M_\odot$ and $\sigma = 0.69$.

These templates are provided in rest-frame air wavelengths with a spectral resolution (FWHM) of 3 Å. We converted the wavelength from air to vacuum using the relation provided by Morton (2000). We then transformed the wavelength from rest-frame to the observed frame using the relation $\lambda_{\rm obs} = \lambda_{\rm rest}(z+1)$.

The GALAXEV software provides templates in units of total wavelength-dependent luminosity, $L(\lambda)$, per unit of total stellar mass (in Solar masses $M_\odot$). To obtain the flux in erg s$^{-1}$ cm$^{-2}$ Å$^{-1}$, we multiplied the luminosity by the total stellar mass, and converted the luminosity into flux using the luminosity distance function, $d_L(z)$, such that,

$$F_{\rm int}(\lambda)[\text{erg s}^{-1}\,\text{cm}^{-2}\,\text{Å}^{-1}] = \frac{L(\lambda)\,M_\star}{4\pi d_L^2(z)\,(1+z)}. \quad (5)$$

The total stellar mass ($M_\star$) and the redshift ($z$) are those provided in the L16 and B19 catalogues. The $(1+z)$ term in the denominator accounts for the fact that the flux and luminosity are not bolometric but are densities per unit wavelength (Hogg et al. 2002). We finally applied the Calzetti et al. (2000) extinction law to convert the intrinsic flux ($F_{\rm int}$) derived in Eq. (5) into the observed, extinction-corrected flux ($F_{\rm obs}$), as

$$F_{\rm obs}(\lambda) = F_{\rm int}(\lambda)\,10^{-0.4\,A_\lambda}, \quad (6)$$

where the dust extinction is taken into account with the following:

$$A_\lambda = k_\lambda\,E_{\rm star}(B-V). \quad (7)$$

Here $E_{\rm star}(B-V)$ is the colour excess taken from the L16 and B19 catalogues and $k_\lambda$ is the wavelength-dependent extinction curve used by Calzetti et al. (2000), with a normalisation factor of $R_V = 4.05$.

### 3.2. Prediction of the emission lines fluxes

We present in this section the prediction of fluxes of the photoionised emission lines H$\alpha$, H$\beta$, H$\gamma$, H$\delta$, H$\epsilon$, H8, H9, H10, H11, H12 (Balmer lines), P$\beta$, P$\gamma$, P$\delta$, P8, P9, P10 (Paschen lines), and of the collision excited forbidden lines [N II]$\lambda$6584, [N II]$\lambda$6549, [O III]$\lambda$5008, [O III]$\lambda$4959, [O II]$\lambda\lambda$3727,3729, [S II]$\lambda$6731, [S II]$\lambda$6717, [S III]$\lambda$9531, [S III]$\lambda$9069. To determine the emission line fluxes, we referred to the broad-band SED-fitting parameters $M_\star$, SFR, $z$ and $A_V$ available in the L16 and B19 catalogues and used empirical and theoretical relations available in the literature together with their corresponding observed scatters described below.

We anticipate here that we focussed the analysis of the extracted spectra on the H$\alpha$, [O III]$\lambda$5008, and [S III]$\lambda$9531 emission lines. The procedure for predicting the fluxes of the other emission lines is however presented in this section to provide a full description of the construction of the spectral library that is available upon request.

#### 3.2.1. Prediction of the H$\alpha$ line fluxes

The study of the SFR is of particular interest to trace the SFH and gives vital clues of the physical nature of the *Hubble* sequence and evolutionary histories of galaxies (Roberts 1969; Larson & Tinsley 1977; Kennicutt 1998; Daddi et al. 2004, 2007; Rodighiero et al. 2011, 2014). The SFR can be inferred from measurements of the integrated light in the UV (see Donas & Deharveng 1984) and far-infrared (see Rieke & Lobofsky 1979), using SED-fitting methods or by tracking and measuring nebular recombination line fluxes (Madau & Dickinson 2014). SFR estimates in the literature obtained from different calibrations and in different redshift bins show consistent results. For example, using multi-wavelength data from the GOODS-N field, Daddi et al. (2007) have found good agreement between SFRs calculated from radio, far-IR, mid-IR, and even UV (corrected for the extinction by dust). More recently, this consistency has been confirmed by Sanders et al. (2020) comparing the SFRs obtained from [O II] observed flux (SFR([O II])) to SFR(H$\alpha$), and by Kashino et al. (2013) comparing the SFR(UV) with SFR(H$\alpha$).

In this study we used the SFR derived from SED-fitting available in the L16 and B19 catalogues.

Kennicutt (1998) presented a calibration between the intrinsic H$\alpha$ luminosity – $L$(H$\alpha$) – and SFR from which we based our flux calculations. Kennicutt (1998) used a Salpeter (1955) IMF while we have adopted the Chabrier (2003) IMF. To account for this difference, we applied a correction dividing by a factor 1.7 to transform the calibration for the use of Chabrier (2003) IMF as presented by Kashino et al. (2019). We then obtained the relation,

$$L(\text{H}\alpha)[\text{erg s}^{-1}] = \frac{\text{SFR}[M_\odot \, \text{yr}^{-1}]}{4.6 \times 10^{-42}}, \quad (8)$$

where SFR[$M_\odot$ yr$^{-1}$] takes the value retrieved from the SED-fitting available in the L16 and B19 catalogues. We converted the luminosity obtained from the Eq. (8) into the intrinsic predicted flux for H$\alpha$ as follows:

$$\text{H}\alpha[\text{erg s}^{-1} \, \text{cm}^{-2}] = \frac{L(\text{H}\alpha)[\text{erg s}^{-1}]}{4\pi d_L^2(z)}. \quad (9)$$

#### 3.2.2. Prediction of the [O II]$\lambda\lambda$3727,3729 line fluxes

The strongest emission feature in the wavelength range 0.35–0.45 μm is the [O II]$\lambda\lambda$3727,3729 forbidden-line doublet which is extremely useful for studies of distant galaxies because it can be observed in the visible out to redshift $z \sim 1.6$. The luminosities of forbidden lines are not directly coupled to the ionising luminosity, and their excitation is sensitive to the abundance and ionisation state of the gas. However, the excitation of [O II] is sufficiently well behaved that it can be calibrated empirically as a quantitative SFR tracer (Kennicutt 1998; Kewley et al. 2004). Even if this calibration suffers from a dependence on secondary parameters, such as the metal abundance, it remains to first approximation a reliable tracer of the current SFR (Kewley et al. 2004). We used the calibration presented by Kewley et al. (2004), equation 4, that has been derived from two samples of galaxies located at high redshift, one at $0.8 \leq z \leq 1.6$, from the NICMOS H$\alpha$ survey, and the other at $0.5 \leq z \leq 1.1$, from the Canada-France Redshift Survey. The calibration is presented as,

$$L([\text{O II}])[\text{erg s}^{-1}] = \frac{\text{SFR}([\text{O II}])[M_\odot \, \text{yr}^{-1}]}{6.58 \times 10^{-42}}, \quad (10)$$

where SFR([O II]) is in agreement with the SFR obtained from H$\alpha$ (Kewley et al. 2004, Eq. (5)). We therefore used the same SFR that we have used to calculate $L$(H$\alpha$), which is the SFR obtained from the SED-fitting.

#### 3.2.3. Prediction of the Balmer and Paschen line fluxes

We predicted H$\beta$ and the other Balmer (H12) and Paschen (P10) line fluxes assuming ratios between their respective intrinsic fluxes starting from the H$\alpha$ fluxes predicted above. We made use of the ratios presented by Hummer & Storey (1987) and Osterbrock (1989). These lines theoretically also directly trace the SFR, but being relatively faint, they are rather difficult to detect compared to the H$\alpha$ line, and other forbidden lines, for example [O II] and [O III]. However, if they are detected, they can be used to improve the redshift measurement precision.

The hydrogen Balmer decrement, H$\alpha$/H$\beta$, is frequently used to determine the amount of dust extinction for the low-density gas component by comparing the observed H$\alpha$/H$\beta$ ratio with the expected intrinsic ratio. An intrinsic value of 2.86 is generally adopted assuming typical H II region gas conditions where the electron density is $n_e = 10^2$ cm$^{-3}$ with an electron temperature $T_e = 10^4$ K and assuming case B recombination (see Osterbrock 1989). We considered this ratio to estimate the intrinsic Hbeta flux of our galaxies.

From the intrinsic predicted flux for H$\beta$ and referring to Hummer & Storey (1987), we predicted the intrinsic fluxes for all the other Balmer and the Paschen lines. These lines are namely the H$\gamma$, H$\delta$, H$\epsilon$, H8, H9, H10, H11, H12 (Balmer), P$\beta$, P$\gamma$, P$\delta$, P8, P9, P10 (Paschen) lines (see Table 1).

#### 3.2.4. Prediction of the [N II]$\lambda\lambda$6584,6549 doublet fluxes

The strongest line of the [N II] doublet, [N II]$\lambda$6584, can be very useful in increasing the confidence in redshift determination (Silverman et al. 2015), in estimating the gas-phase metallicity, and even in yielding insight on the kinematics from the FWHM of the line. It has therefore been extensively studied, usually by measuring observed ratio [N II]$\lambda$6584/H$\alpha$, often mentioned in the literature as $N_2$. The $N_2$ ratio is sensitive to the metallicity which is measured by the oxygen abundance (O/H; Denicolo et al. 2002; Henry et al. 2001). So far, it has been possible to study H$\alpha$ and [N II] in thousands of galaxies up to $z \sim 2.5$ (reaching the $K$ band wavelength limit) such that the $N_2$ index offers a unique way to study the evolution of the metallicity with redshift (Steidel et al. 2014; Pettini & Pagel 2004).

**Table 1.** Emission lines in vacuum that we added to the compiled continuum SED.

| | Emission lines | Wavelength (Å) | Ratio H(−)/H$\beta$ |
|---|---|---|---|
| Balmer lines | H$\alpha$ | 6564.61 | 2.86 |
| | H$\beta$ | 4862.69 | 1 |
| | H$\gamma$ | 4341.69 | 4.68×10$^{-1}$ |
| | H$\delta$ | 4102.92 | 2.59×10$^{-1}$ |
| | H$\epsilon$ | 3971.19 | 1.59×10$^{-1}$ |
| | H8 | 3890.15 | 1.05×10$^{-1}$ |
| | H9 | 3836.48 | 7.31×10$^{-2}$ |
| | H10 | 3798.98 | 5.30×10$^{-2}$ |
| | H11 | 3771.70 | 3.97×10$^{-2}$ |
| | H12 | 3751.22 | 3.05×10$^{-2}$ |
| Paschen Lines | P$\beta$ | 12 821.59 | 1.63×10$^{-1}$ |
| | P$\gamma$ | 10 941.09 | 9.04×10$^{-2}$ |
| | P$\delta$ | 10 052.13 | 5.55×10$^{-2}$ |
| | P8 | 9548.59 | 3.66×10$^{-2}$ |
| | P9 | 9231.55 | 2.54×10$^{-2}$ |
| | P10 | 9017.39 | 1.84×10$^{-2}$ |
| [N II] | [N II]$\lambda$6584 | 6585.23 | − |
| | [N II]$\lambda$6549 | 6549.84 | − |
| [O III] | [O III]$\lambda$5008 | 5008.24 | − |
| | [O III]$\lambda$4959 | 4960.30 | − |
| [O II] | [O II]$\lambda\lambda$3727,3729 | 3728.49 | − |
| [S II] | [S II]$\lambda$6731 | 6732.71 | − |
| | [S II]$\lambda$6717 | 6718.32 | − |
| [S III] | [S III]$\lambda$9531 | 9533.20 | − |
| | [S III]$\lambda$9069 | 9071.10 | − |

**Notes.** We predicted the fluxes for these emission lines using the relations presented in Sect. 3.2. The rightmost column lists the ratios presented by Hummer & Storey (1987) and Osterbrock (1989) between the fluxes of the Balmer and Paschen lines and flux of the H$\beta$ line assuming typical H II region gas conditions with electron density $n_e = 10^2$ cm$^{-3}$, electron temperature $T_e = 10^4$ K, and case B recombination.

Given the dependence of the [NII] doublet lines on metallicity, we needed an estimate of this physical quantity to derive the expected [N II] doublet flux. In particular, we made use of the $M_\star$-metallicity relation (MZR) derived empirically by Wuyts et al. (2014), based on the observations of a sample of 222 galaxies at $0.8 \leq z \leq 2.6$ and $9.0 \leq \log_{10}(M_\star[M_\odot]) \leq 11.5$ (see Wuyts et al. 2014, for further details). Based on the parameterisation originally proposed by Zahid et al. (2014), we referred to Wuyts et al. (2014), Eq. (1), for the redshift-dependent MZR. Wuyts et al. (2014) also showed that their MZR parameterisation is in good agreement with that obtained by Zahid et al. (2014) at $0 \leq z \leq 1.6$, as well as with results for other high-$z$ samples at $2.2 \leq z \leq 2.3$ (Erb et al. 2006; Steidel et al. 2014).

The oxygen abundance $12 + \log_{10}(O/H)$ obtained from Eq. (1) in Wuyts et al. (2014) should then be converted into a value for the ratio $N_2$. For this purpose, we made use of the linear metallicity calibration proposed by Pettini & Pagel (2004) based on 137 extragalactic H II regions with well determined values of (O/H) and $N_2$. The linear calibration is then presented as follows:

$$12 + \log_{10}(O/H) = 8.90 + 0.57\, N_2. \quad (11)$$

The scatter of this relation for local galaxies is equal to 0.18 dex.

However, to reproduce a plausible distribution of higher redshift galaxies in the $N_2$-$M_\star$ plane, we used the intrinsic scatter on $N_2$ inferred by Kashino et al. (2019) on the FMOS-COSMOS survey sample at $1.43 \leq z \leq 1.74$, which has been shown to be very consistent with that derived for a sample of SDSS local galaxies. The intrinsic scatter $\sigma_{\rm int}(N_2)$ is derived as a function of $N_2(M_\star)$, that is the best-fit $M_\star$–$N_2$ relation and is presented in Eq. (20) of their paper. The results from this approach are presented in Fig. 4. From this predicted $N_2$ ratio, we obtained the flux for the emission line [N II]$\lambda$6584 using the previously calculated flux of the H$\alpha$ line.

We inferred the flux of the second line of the [N II] doublet, namely [N II]$\lambda$6549, using the relation [N II]$\lambda$6584/[N II]$\lambda$6549 = 2.95 as proposed by Acker et al. (1989). We anticipate here that the [N II] doublet will be blended with the H$\alpha$ line in the *Euclid* extracted spectra (see Sect. 5).

### 3.2.5. Prediction of the [O III]$\lambda\lambda$4959,5008 fluxes

The Baldwin-Phillips-Terlevich (BPT) diagram, introduced by Baldwin et al. (1981), is a powerful tool to infer physical properties of emission-line galaxies.

The BPT diagram of $N_2$ versus $\log_{10}([\text{O III}]\lambda5008/\text{H}\beta)$ (hereafter $O_3$), known as the $N_2$-BPT diagram, happens to be a particularly efficient tool in discriminating the dominant excitation mechanism of nebular emission in galaxies, providing a clear separation of galaxies whose spectra are dominated by ionisation induced by the UV radiation field of young stars from those essentially ionised by the extreme ultra-violet (EUV) radiation of AGNs. This interesting feature led astronomers to extensively refer to the BPT diagram for AGN classification schemes based on observations and photoionisation models, stellar population synthesis and shock modelling (Osterbrock & Podge 1985; Veilleux & Osterbrock 1987; Kewley et al. 2001, 2013a). Of interest for this work is the fact that SFGs form a tight sequence in the $N_2$-$O_3$ plane. The study of the emission line [O III]$\lambda$5008 is not only useful to identify the presence of an AGN (Kewley et al. 2013a,b; Kartaltepe et al. 2015), but it is also crucial to determine the oxygen enrichment of the interstellar medium (ISM; Zahid et al. 2014).

As described in Sects. 2.1 and 2.2, we restricted our sample to be exclusively composed of normal SFGs without AGNs. Making this assumption enabled us to make use of the calibration established for the SFG abundance sequence on the BPT diagram (Kewley et al. 2013a).

Thus, using the $N_2$ ratio calculated in Sect. 3.2.4, we could predict the $O_3$ ratio. The calibration provided by Kewley et al. (2013a) for galaxies located at $0 \leq z \leq 3$, started from a previous study of local SDSS galaxies (Kewley et al. 2006) and extended towards higher redshift using the chemical evolution predictions from cosmological hydrodynamic simulations with theoretical stellar population synthesis, photoionisation and shock models (see Kewley et al. 2013a, Fig. 3).

The calibration of the SFGs abundance sequence in the $N_2$-BPT diagram proposed by Kewley et al. (2013a) is presented as follows:

$$O_3 = 1.2 + 0.03\, z + \frac{0.61}{N_2 - 0.02 - 0.1833\, z}. \quad (12)$$

This calibration is scattered along both the $x$ and $y$ axes by 0.1 dex. For our purpose, we fixed the value of $N_2$, since a realistic scatter on $N_2$ was already introduced in Sect. 3.2.4, and therefore we only kept the vertical 0.1 dex scatter that we distributed normally. We present in Fig. 5 the predicted $N_2$-BPT diagrams

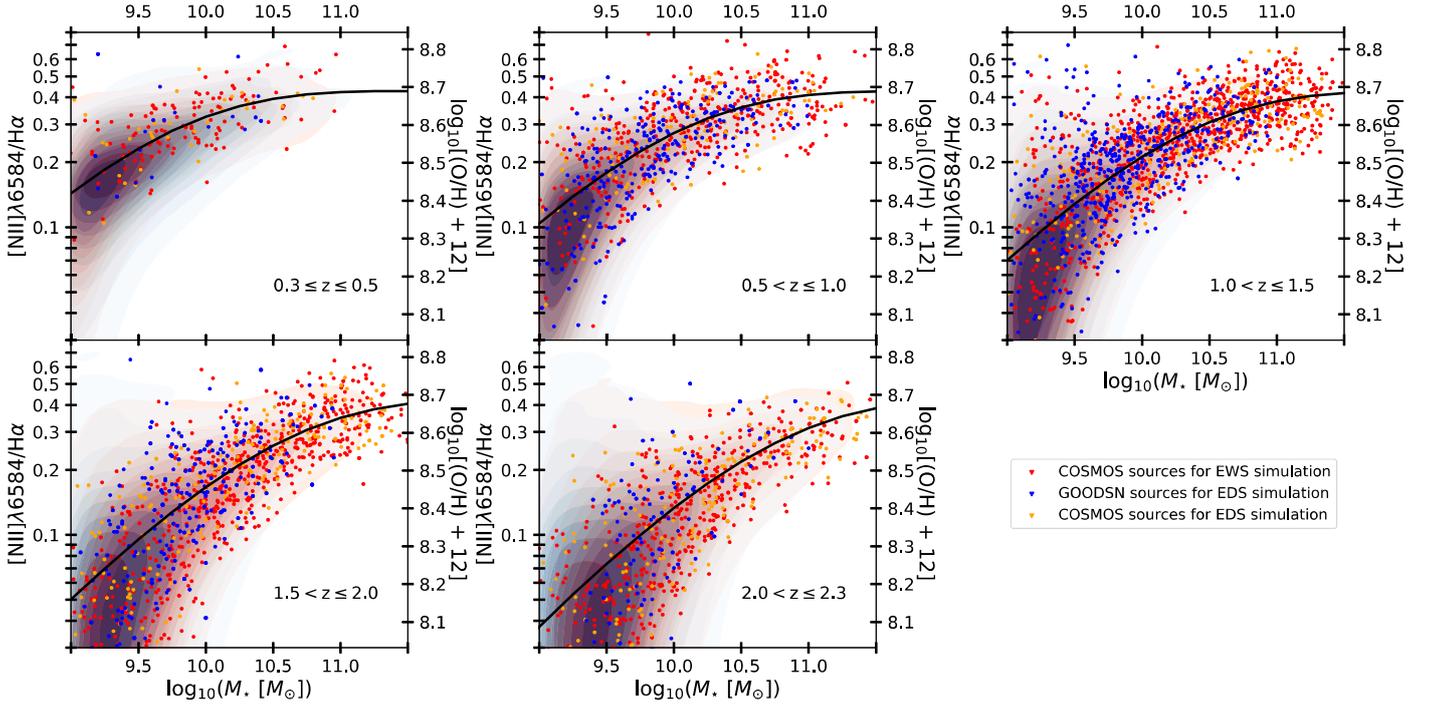

**Fig. 4.** $M_\star$-metallicity relation (MZR) in five redshift bins. The solid black lines show the redshift-dependent MZR proposed by Wuyts et al. (2014) in their Eqs. (1) and (2). The predicted metallicity is indicated on the right vertical axis of each plot. The left vertical axis indicates the corresponding [N II]$\lambda 6584$/H$\alpha$ flux ratio predicted using the linear calibrations presented in Asplund et al. (2004). Readers can refer to the caption of Fig. 1 for a colour assignment description.

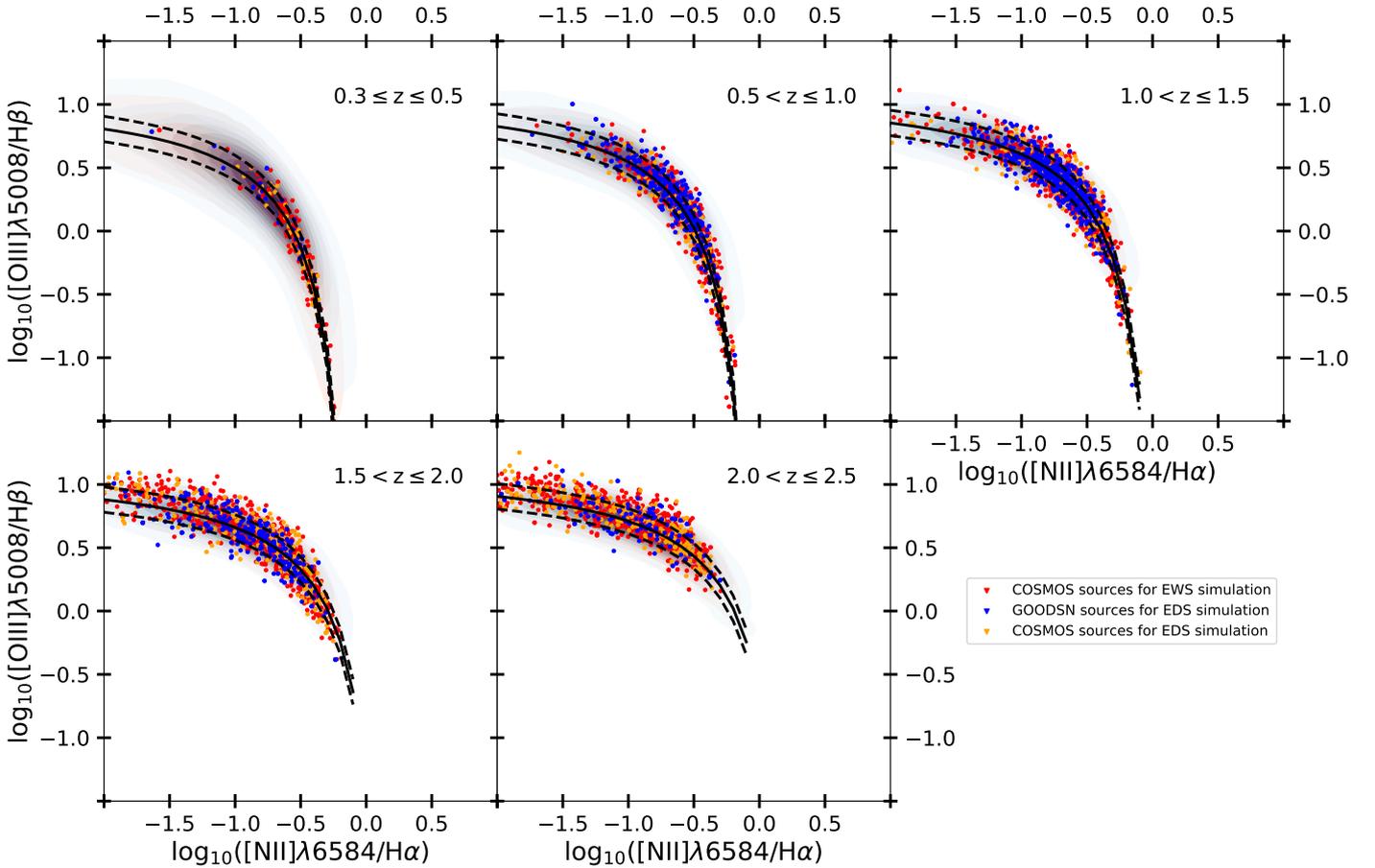

**Fig. 5.** $N_2$-BPT diagrams in five redshift bins. The solid black line shows the relation derived by Kewley et al. (2013a). The dashed lines represent the scatter that has been applied 'vertically' by fixing the ratio $N_2$ and applying a scatter of 0.1 dex on our predicted ratio $O_3$. Readers can refer to the caption of Fig. 1 for a colour assignment description.

in different redshift ranges for sources from the L16 the B19 catalogues.

From this predicted $O_3$ ratio, we obtained the flux for the emission line [O III]$\lambda$5008 using the previously calculated H$\beta$ flux. We inferred the flux of the second line of the [O III] doublet, [O III]$\lambda$4959, using the theoretical ratio presented by Storey & Zeippen (2000) such that [O III]$\lambda$5008/[O III]$\lambda$4959 = 2.98.

### 3.2.6. Prediction of the [S II]$\lambda\lambda$6717,6731 fluxes

An alternative formulation of the BPT diagram compares the flux ratios [S II]$\lambda\lambda$6717,6731/H$\alpha$ and [O III]$\lambda$5008/H$\alpha$. This BPT diagram can be used to discriminate SFGs from AGNs (Veilleux & Osterbrock 1987; Dopita et al. 2016; Kashino et al. 2017) even though the separation is not as clear as for the $N_2$-BPT diagram presented above.

We made use of the calibration between these ratios derived for SFGs, which has been introduced for local sources by Dopita et al. (2016) and confirmed at high-redshift by Kashino et al. (2017, see Fig. 7 of their paper), on a sample of 701 SFGs with stellar masses $9.6 \leq \log_{10}(M_\star[M_\odot]) \leq 11.6$, located at $1.4 \leq z \leq 1.7$. The calibration is presented as,

$$12 + \log_{10}(\mathrm{O/H}) = 8.77 + N_2S_2 + 0.264\,N_2, \tag{13}$$

where $N_2S_2 = \log_{10}([\mathrm{N\,II}]\lambda6584/[\mathrm{S\,II}]\lambda\lambda6717,6731)$ and $\log_{10}(\mathrm{O/H})$ is the oxygen abundance.

For the transformation of $N_2$ into the oxygen abundance, we used the linear calibration proposed by Pettini & Pagel (2004) while Kashino et al. (2017) made use of the calibration by Tremonti et al. (2004). We then referred to Kewley & Ellison (2008, Eq. (1) and Table 3) to convert metallicity relations into any other calibration scheme. It is formulated for our purpose as,

$$\begin{aligned} y = &-1661.9380 + 585.17650\,x \\ &- 68.471750\,x^2 + 2.6766690\,x^3, \end{aligned} \tag{14}$$

where $y$ is the metallicity in $12 + \log_{10}(\mathrm{O/H})$ units as calibrated by Tremonti et al. (2004), and $x$ is the original metallicity (Pettini & Pagel 2004) to be converted, also in $12 + \log_{10}(\mathrm{O/H})$ units. We added a scatter with a normal distribution with sigma 0.1 dex on the $N_2S_2$ value calculated from Eq. (13), from which we eventually obtained the sum of the fluxes of the [S II] doublet. The 0.1 dex has been added to implement a scatter corresponding to the median scatter of the $N_2S_2$ distribution on the BPT diagrams presented by Dopita et al. (2016) and Kashino et al. (2017).

For the determination of the ratio between the lines of the doublet [S II]$\lambda\lambda$6717, 6731, we relied on the relation between the electron density ($n_e$) and the ratio of the fluxes of the two forbidden lines of interest. Indeed, for a pair of lines with nearly the same excitation state, as can be an ion excited in two different states in a medium with typical low density such as in an H II region and if both excitation states are metastable, a relation can be established between the electron density and the ratio of the lines [S II]$\lambda$6716/[S II]$\lambda$6731. A relation has been derived by Proxauf et al. (2014) following previous research made by Osterbrock & Ferland (2006) and is presented in Iani et al. (2019), Eq. (2).

To obtain $n_e$, we referred to a relation between $n_e$ and the sSFR proposed by Kashino & Inoue (2018), Eq. (10). They obtained this relation from a sample of galaxies located at $0.027 \leq z \leq 0.25$. They estimated $n_e$ from the ratio [S II]$\lambda$6717/[S II]$\lambda$6731, assuming the electron temperature $T_e(\mathrm{S}_+) = T_e(\mathrm{O}_+)$ where $T_e(\mathrm{O}_+)$ is estimated from the ratio [O II]$\lambda\lambda$3726,3729/[O II]$\lambda\lambda$7320,7330. A trend for the ratio [S II]$\lambda$6717/[S II]$\lambda$6731 as a function of the electron density is presented in Iani et al. (2019), Fig. 9.

### 3.2.7. Prediction of the [S III]$\lambda\lambda$9069,9531 fluxes

The doublet [S III]$\lambda\lambda$9069,9531 has been challenging to study due to its relatively long wavelength, lying at the edge of classical optical spectrometers even for nearby galaxies. The doublet will fall in the RGS passband up to a redshift $z \sim 0.9$. [S III]$\lambda\lambda$9069,9531 is generally studied through the strong-line ionisation parameter diagnostic $S_{32}$ which corresponds to the ratio ([S III]$\lambda$9069 + [S III]$\lambda$9531)/[S II]$\lambda\lambda$6717,6731. This ratio was introduced by Kewley & Dopita (2002) and has been referred to in more recent studies (Morisset et al. 2016; Sanders et al. 2020; Kewley et al. 2019).

In this study, we referred to a relation presented by Kewley et al. (2019). They have shown that the $S_{32}$ ratio is sensitive to the ionisation parameter ($U$), with a relatively small variation lower than 0.3 dex in our range of metallicity $8.1 \leq 12 + \log_{10}(\mathrm{O/H}) \leq 8.7$ (see Fig. 4). This calibration still suffers from a complex temperature dependence that can make the photoionisation models underestimate the [S II] line strength (Levesque et al. 2010; Kewley et al. 2019) and should therefore be considered with caution. We calculated the $\log_{10}(U)$ using the relations presented by Kashino & Inoue (2018), Eqs. (11) and (12), for consistency with the previously calculated electron density ($n_e$).

We inferred a best linear fit to the results presented by Kewley et al. (2019) as,

$$R = 0.75 \log_{10}(U) + 2.625, \tag{15}$$

with,

$$R = \frac{[\mathrm{S\,III}]\lambda 9069 + [\mathrm{S\,III}]\lambda 9531}{[\mathrm{S\,II}]\lambda\lambda 6717, 6731}. \tag{16}$$

Finally, we inferred the flux of the individual lines of the doublet using the value for the theoretical ratio [S III]$\lambda$9531/[S III]$\lambda$9069 = 2.5 as proposed by Sanders et al. (2020).

### 3.2.8. From intrinsic to observed emission line fluxes

For the reddening, we proceeded in a similar way as we did for the continuum in Sect. 3.1, although accounting for the difference between the $E_{\mathrm{star}}(B-V)$ of the stellar component and the $E_{\mathrm{neb}}(B-V)$ of the line-emitting nebular region with a proportionality defined by $f = E_{\mathrm{star}}(B-V)/E_{\mathrm{neb}}(B-V)$. The physical meaning of the $f$-factor is still a subject of debate. Puglisi et al. (2016) have suggested that the $f$-factor is a function of the mass and SFR of the galaxy while Rodríguez-Muñoz et al. (2022) highlighted a stronger correlation with the UV attenuation. Different values have been inferred ranging from 0.44 for local galaxies (Calzetti et al. 2000) up to ~1 in other studies at higher redshifts (Kashino et al. 2013; Puglisi et al. 2016; Rodríguez-Muñoz et al. 2022). Some of the differences among these studies are to be ascribed to the different galaxy samples analysed, for example normal SFGs versus heavily obscured starburst galaxies. In our study we used the value of 0.58[6] for the $f$-factor derived by Calzetti et al. (2000) for local galaxies and proved to be a fair estimate also for high-redshift galaxies (e.g. Förster Schreiber et al. 2009). The observed flux is derived from the intrinsic flux using the Eq. (6) with $A_\lambda$ defined as,

$$A_\lambda = k_\lambda \frac{E_{\text{star}}(B-V)}{f}. \qquad (17)$$

---

[6] We note that the value 0.58 is obtained when applying the Calzetti reddening curve for both the nebular and continuum extinction, while 0.44 is obtained when using the Fitzpatrick reddening curve for the continuum (cf., Puglisi et al. 2016). In our case we apply the Calzetti reddening curve for both the continuum and nebular continuum. We thus chose 0.58 for the $f$-factor.

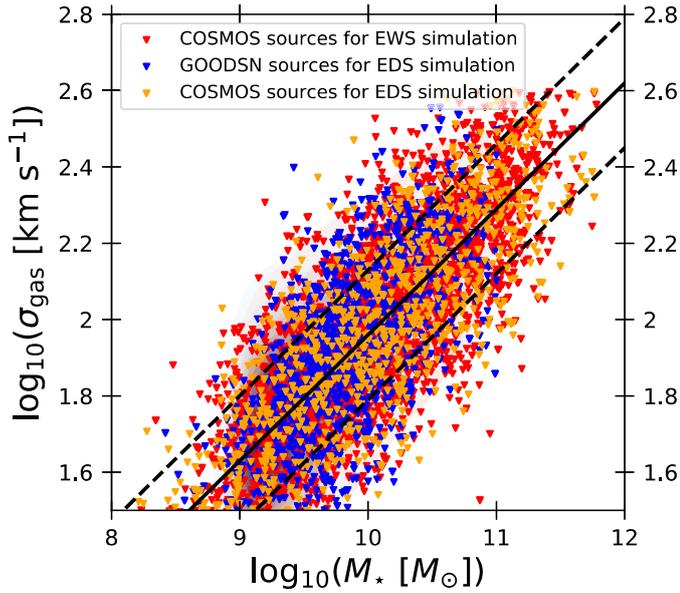

**Fig. 6.** Predicted velocity dispersion in km s$^{-1}$ as a function of the total stellar mass using the relation from Bezanson et al. (2018), indicated with the solid black line. The dashed black lines indicate the 0.17 dex scatter that we applied as explained in the text. Readers can refer to the caption of Fig. 1 for a colour assignment description.

As for the continuum, we obtained the factor $k_\lambda$ using the wavelength-dependent extinction law provided by Calzetti et al. (2000) and taking $R_V = 4.05$. $E_{\text{star}}(B-V)$ is available in the L16 and B19 catalogues.

### 3.3. Integration of the emission lines to the continuum

The predicted fluxes are transformed into lines with a Gaussian profile applying a dispersion in the rest-frame of 3 Å as provided by the models generated through GALAXEV (see Sect. 3.1). We therefore redshifted the dispersion by applying the usual $(1 + z)$ factor such that $\Delta\lambda_{\text{obs}} = \Delta\lambda_{\text{rest}}(1 + z)$. The resulting emission line SED profiles were then added to the continuum SED of each galaxy. We broadened the resulting galaxy SED using the recipe correlating the velocity dispersion of the ionised gas ($\sigma_{\text{gas}}$) to the total stellar mass ($M_{\text{gas}}$) presented in Eq. (3) of Bezanson et al. (2018), which was obtained from the observation of about 1000 massive galaxies with the VLT/VIMOS in the LEGA-C survey. This relation has a scatter of 0.17 dex and has been derived for galaxies at $0.6 \leq z \leq 1.0$. It has also been confirmed for higher redshift $z \sim 2$ using data from the SINS/zC-SINF survey by Förster Schreiber et al. (2018). See Fig. 6 for the results.

Once the $\sigma_{\text{gas}}$ is determined, we broadened the galaxy SED by convolving the spectra with a Gaussian kernel. To cope with the $\sigma_{\text{gas}}$ dependence on the wavelength, we made use of the penalised pixel-fitting method proposed by Cappellari (2017).

### 3.4. Evaluation of the incident galaxy SEDs

#### 3.4.1. Reconstruction of the continuum

We compared the reconstructed continuum of the incident spectra with photometric data found in the literature and in particular data from the Survey for High-$z$ Absorption Red and Dead Sources (SHARDS; Pérez-González et al. 2013) for sources from the B19 catalogue. The SHARDS, part of the CANDELS survey (Grogin et al. 2011; Koekemoer et al. 2011), observed the entire GOODS-N field in 25 medium-band filters with GTC/OSIRIS in the wavelength range 0.5–0.95 µm with contiguous passbands. We also referred for comparison to HST data taken with its Wide Field Camera 3 (WFC3) and Advanced Camera for Surveys (ACS) for the incident spectra built from both L16 and B19 catalogues.

We present in Fig. 7 an example of the very good agreement coming from the photometric check using SHARDS data for B19 sources and HST data for L16 sources data. We also indicate on the figure the predictions in flux measurement using the *Euclid* $I_E$, $Y_E$, $J_E$, and $H_E$ filters. In Fig. 8, we present examples of incident spectra with different redshift and stellar mass.

#### 3.4.2. Evaluation and calibration of the emission line fluxes using spectroscopic data

Using sources that have available emission line fluxes observed in public spectroscopic surveys, we provide a comparison between our predicted fluxes with observations. The results are presented in Fig. 9.

We used data obtained from the LEGA-C, FMOS and 3D-HST surveys. 3D-HST and FMOS do not have sufficient resolution to deblend H$\alpha$ and [N II]$\lambda\lambda$6584,6549 (hereafter the triplet H$\alpha$-[N II]$\lambda\lambda$6584,6549 is referred as H$\alpha$-complex), or the [O III] doublet. For comparison with data from these surveys, we therefore summed our predicted fluxes of the blended lines accordingly. To make our calculated fluxes be as close as possible to observations, we had to apply a factor of 1.8 and 2 to the predicted [O II]$\lambda\lambda$3726,3729 and [S II]$\lambda$6731 fluxes respectively. The reason for these offsets is still unclear due to a relatively poor coverage by current spectroscopic surveys in the redshift range of interest. We present in Fig. 9 the comparison including these correction factors.

## 4. Pilot run simulations: Frame and setup

The pilot run, which is part of the *Euclid* legacy science efforts, aims to provide the *Euclid* community with the following: i) simulations of the spectroscopic channel using the NISP simulator (SIM/TIPS[7]; Zoubian et al. 2014), which are processed through the spectroscopic image reduction (SIR) pipeline[8]; and ii) preliminary results on the spectroscopic capabilities of the RGS for the EWS and EDS. Both the SIM and SIR processing functions are undergoing continuous improvements to add features to the simulations and to optimise the spectral extraction. Hence, the results presented here provide a baseline, but we expect to see improvements in the processing as development continues.

---
[7] This Is a Pixel Simulator (TIPS) is part of the effort in setting up the science ground segment (SGS) that is responsible to carry out the entire data processing, the cosmological analysis, and to deliver the scientific results of the mission. In particular, TIPS is the official pipeline for simulating the NISP-S images within the organisation unit for simulation (see Serrano et al., in prep, for reference on the pre-launch SGS simulation framework), which is in charge of the simulations to construct and validate the *Euclid* mission.
[8] SIM/TIPS version 5.4 & SIR version 2.8, combined using the SIR_SPECTROSIM_RUNNER software (Paganin 2022).

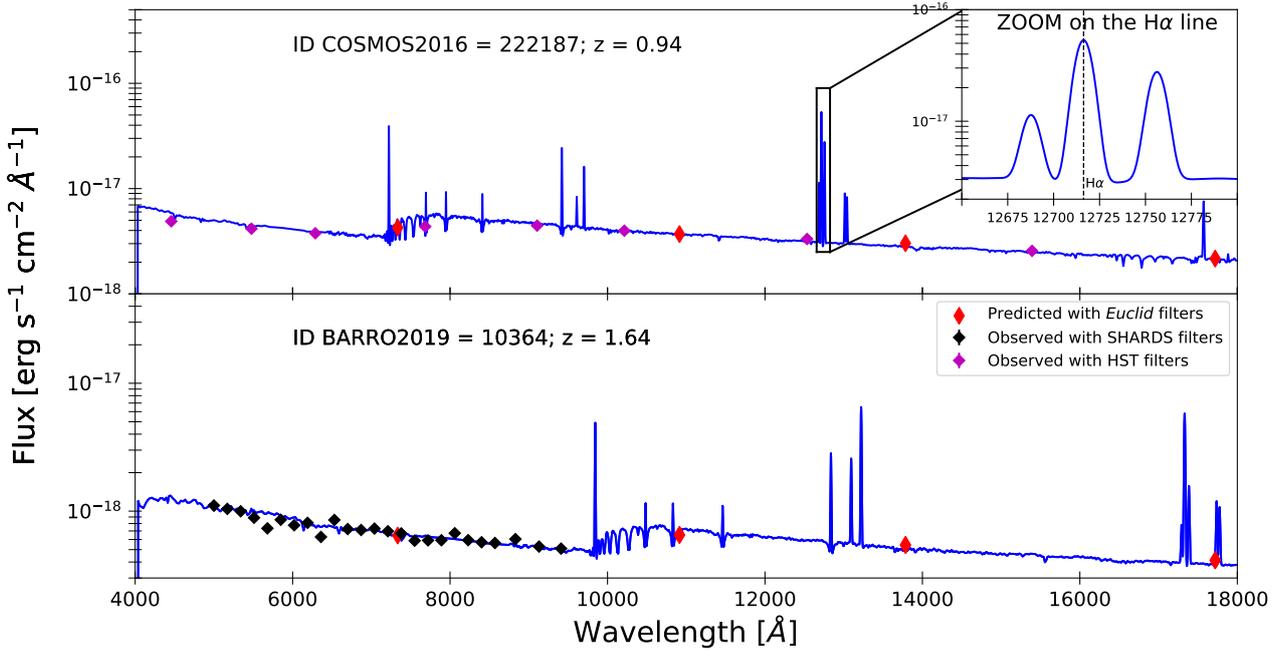

**Fig. 7.** Examples of photometric evaluation of the incident spectra created in this work. *Top*: one incident spectrum from the L16 catalogue compared to observational data from HST photometry (purple diamonds). *Bottom*: one incident spectrum from the B19 catalogue compared to observational data from the SHARDS survey photometry (black diamonds). The blue line shows the spectrum resulting from the combination of the continuum, obtained using the library of evolutionary stellar population synthesis models GALAXEV (Bruzual & Charlot 2003) with our predicted emission lines. *Top right*: zoom on the H$\alpha$-[N II]$\lambda\lambda$6584,6549 that highlights the typical emission line broadening due the velocity dispersion calculated using the relation presented by Bezanson et al. (2018). The red diamonds indicate the integrated fluxes that we obtained by convolving the spectra with the *Euclid* $I_{\rm E}$, $Y_{\rm E}$, $J_{\rm E}$, and $H_{\rm E}$ filters.

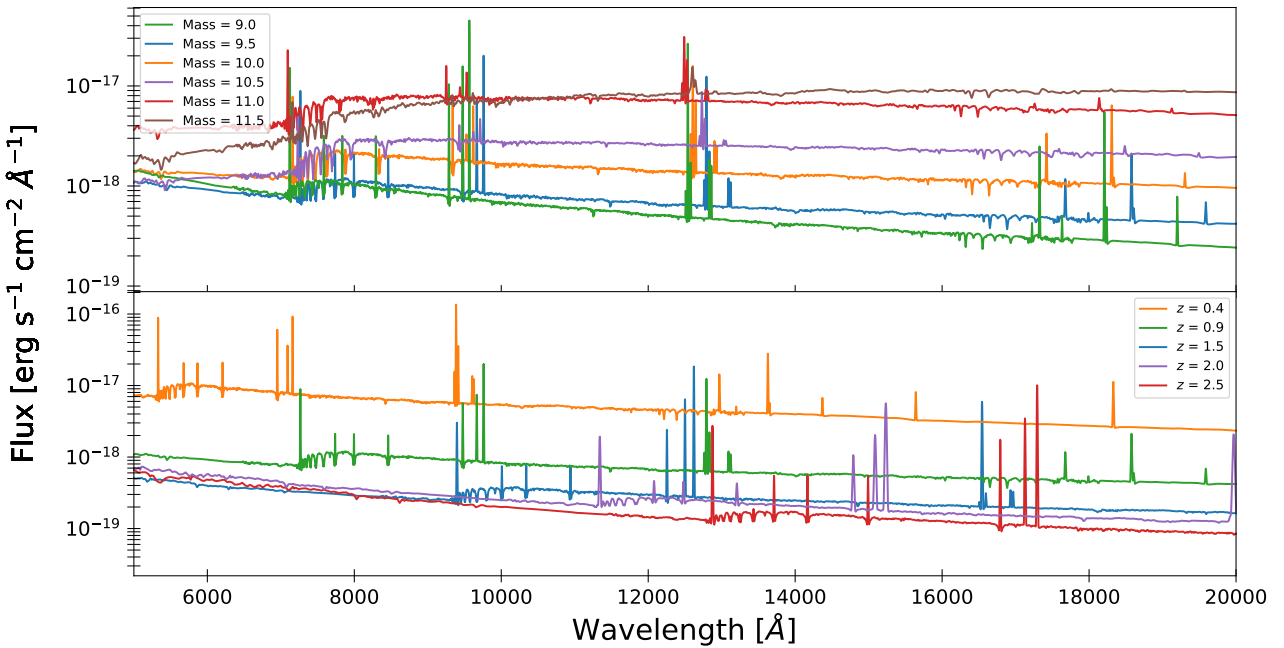

**Fig. 8.** Examples of incident spectra constructed using SED fitting parameters from the L16 catalogue and empirical and theoretical relations for the emission lines. *Top*: spectra of galaxies with different stellar mass, indicated on the figure in $\log_{10}(M_\star\,[M_\odot])$, and at $z = 0.9$. *Bottom*: spectra of galaxies at different redshift and with stellar mass equal to 9.5 in $\log_{10}(M_\star\,[M_\odot])$.

The spectral libraries constructed above have been formatted to be processed through TIPS that simulates the actual RGS observation configurations for the EWS and EDS. The EWS and EDS simulations refer to the corresponding *Euclid* integration time using the RGS. The integration time with the RGS in the EWS simulation is 2212 s which is split into four dithered exposures (each of which is 553.0 s), as defined by the *Euclid* reference observation sequence to cope with pixel defects and cosmic rays. Each dithered exposure was simulated reproducing the up-the-ramp MULTIACCUM acquisition mode for the NISP-S exposures, consisting of 15 groups with 16 readouts per group, and 11 dropped frames (drops) between each group (see Robberto 2014; Kubik et al. 2016, for further details on the acquisition mode). The sensitivity curve to convert the extracted quantity into physical units is presented in Fig. 10. Furthermore, the four dithered exposures are made with four RGS orientations required during the mission for decontamination purpose due to the overlapping of the slitless spectra (Scaramella et al. 2022). The orientation angles are $+0°$, $-4°$, $+180°$, and $+184°$. The integration time in the EDS, that includes both RGS and BGS, will be 40 times longer than that in the EWS. Studies to asses the best configuration in terms of the relative integration time for the two grisms are still ongoing. In this work we assumed the configuration including 30 visits (one visit is made of four dithered frames) with the BGS and 10 visits with the RGS. The EDS integration time with the RGS then is 22 120 s which is split into 40 dithered exposures and will be complemented with 120 dithered exposures using the BGS; however, the latter is not considered in the pilot run since the related pipeline was still undergoing fundamental validation tests when the simulations were performed. The 40 dithered frames of the EDS simulations have been obtained repeating ten times the EWS simulations, with a further 'circular' dither pattern obtained shifting the pointing by 0″.9, that is three pixels, at each step in order to remove artefacts, for example bad pixels, in post-processing.

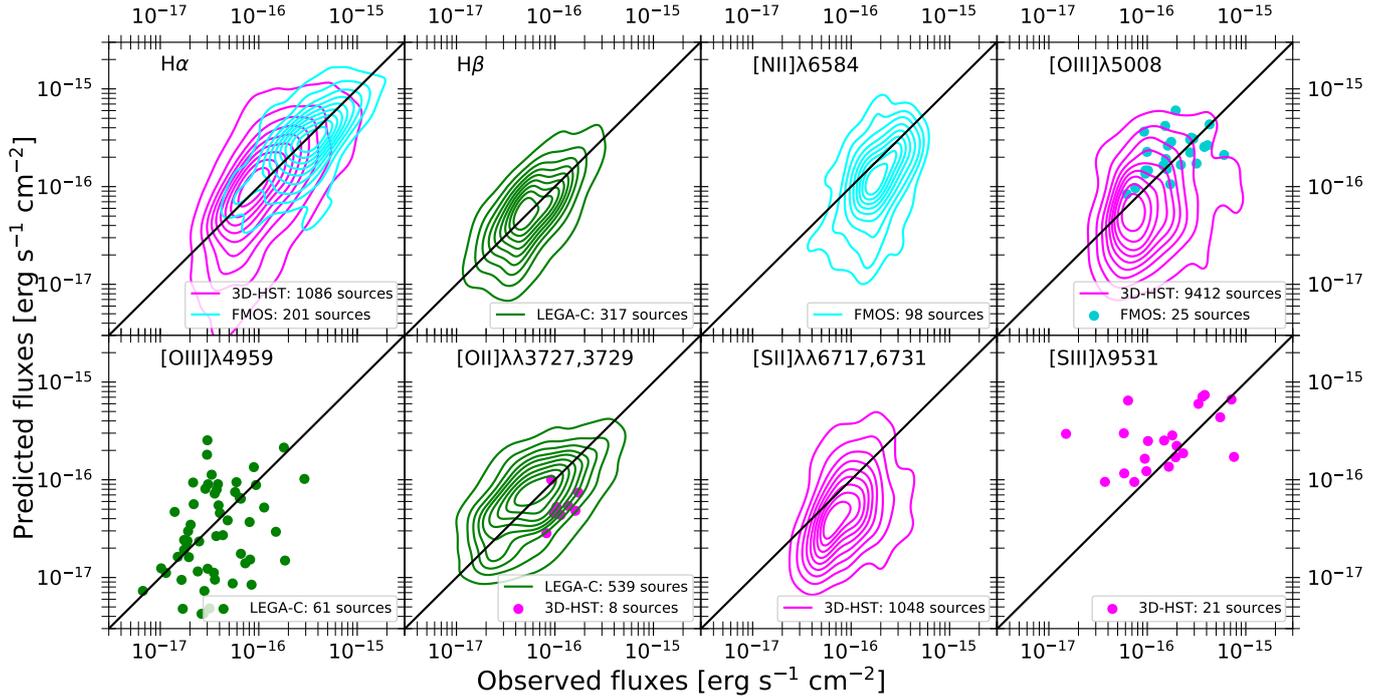

**Fig. 9.** Comparison of the predicted fluxes to the observed fluxes coming from publicly released data of the near-infrared spectroscopic surveys 3D-HST (purple), FMOS (green) and LEGA-C (cyan). The contours correspond to iso-proportions of the distribution density starting at 20% with a 10% step. The scatter plots are indicated for the smallest samples. For our comparison we selected emission line measurements observed with $S/N \geq 3.5$.

In practice, the simulator places each simulated galaxy in a certain position of a simulated sky and then recreates the output of the observations through the instrument. In the simulations performed in this work, we pointed the telescope towards the RA = 228°.394 and Dec = 6°.590 coordinates which determines the background level in the images. The simulations were constructed considering two astrophysical background sources uniformly distributed across the field of view, which come to

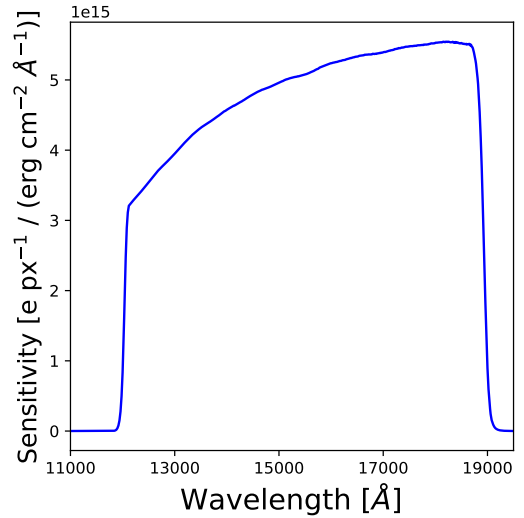

**Fig. 10.** Wavelength-dependent sensitivity curve of the RGS inferred by TIPS using the NISP and payload module transmission curves characterised at nine positions of the NISP focal plane during the ground-test campaigns (Waczynski et al. 2016; Barbier et al. 2018; Costille et al. 2019; Maciaszek et al. 2022) and spatially averaged in the simulations computing the arithmetic mean of the transmissions measured at the nine positions. Units on the vertical axis are in $e\,s^{-1}\,px^{-1}/(erg\,s^{-1}\,cm^{-2}\,\text{Å}^{-1})$. This quantity connects the spectra extracted from the slitless data to the physical units presented in this paper.

dominate the noise level on the detector: the zodiacal light[9] was predicted for the Euclid Survey (Scaramella et al. 2022)

---

[9] The zodiacal light in the EWS is expected to vary between 1.1 and 3.0 photon $s^{-1}\,px^{-1}$, with a median value at 1.6 photon $s^{-1}\,px^{-1}$ and is estimated at 2.2 photon $s^{-1}\,px^{-1}$ at our simulated pointing coordinates.

using the (Aldering 2001) model with an angular dependence by Leinert et al. (1998), and the out-of-field stray light[10] modelled by Venancio et al. (2016, 2020) and presented in Scaramella et al. (2022). In addition to these astrophysical backgrounds, the detector noise has contributions from the readout noise, dark current, and quantum efficiency that were characterised during the NISP ground test campaigns (Waczynski et al. 2016; Barbier et al. 2018; Costille et al. 2019; Maciaszek et al. 2022).

The simulator transforms the galaxy morphological parameters to pixelised galaxy light profiles (Serrano et al, in prep.) based on the GALSIM software (Rowe et al. 2015). The bulge and disk components are computed using two different Sérsic profiles, considering a thick exponential disk model with Sérsic index equal to one (van der Kruit & Searle 1981; Bizyaev 2007). The PSF measurements of the NISP-S, made on the four RGS, in different positions of the focal plane, and at different wavelengths during the ground test campaign, have been parameterised into TIPS. The simulations therefore reproduce the response at the pixel level of the 16 detectors of the NISP focal plane. In a similar way, the spectral dispersion has been measured in laboratory on average at $13.51 \pm 0.06$ Å px$^{-1}$ (W. Gillard et al., in prep) and has been fixed at 13.4 Å px$^{-1}$ for the simulations. The simulation does not include the spectral and astrometric distortions such that the spectra have their nominal inclination and position. The actual inclination and dispersion of the spectra may be affected by mechanical considerations, for example launch vibrations, zero gravity conditions, and thermal stabilisation, that will be verified in orbit during the performance verification test. In particular the spectral dispersion on different positions of the focal plane will be assessed using compact planetary nebulae with known emission lines (Paterson et al. 2023).

As anticipated in Sect. 2.3, the simulated pointings have been populated with a fixed number of 2496 galaxies. This was a choice well motivated by the purpose of first assessment of the RGS capabilities independently from the contamination due to the overlap of the spectra. We avoided the overlap of the 2D spectra on the detector by locating the 2496 galaxies on the field of view following an ordered pattern ensuring that the spectra fell on one single detector avoiding the edges.

## 5. Analysis of the spectra produced by TIPS

We processed the 2D spectra produced by TIPS through the SIR pipeline. SIR is the official pipeline developed for *Euclid* that performs the image reduction, the wavelength and flux calibration, and the extraction of the 1D spectra. The 1D spectra are extracted by SIR using the 'virtual slit' formed by the object itself (see Kümmel et al. 2009). In this work, the SIR pipeline extracted the first order spectra for all the sources of the input catalogue. The spectral extraction uses a fixed aperture in pixels that depends on the shape of the galaxy as described below. SIR processes individual frames and then combines the extracted spectra using inverse-variance weighting. The error of the single frames is propagated assuming a Gaussian behaviour, and so the error on the combined spectra is inversely proportional to the square root of the number of dithered frames. SIR provides the 1D combined spectra and variance that are defined on the wavelength range 1.19–1.90 μm with a pixel scale of 13.4 Å. We proceeded to run the same pipeline routines to extract the 1D spectra from the 40 dithered frames for the EDS simulation and produce the 1D combined spectra products.

---
[10] The stray light during the EWS is expected to vary between 0.1 and 3.0 photon s$^{-1}$ px$^{-1}$, with a median value at 0.4 photon s$^{-1}$ px$^{-1}$ and is estimated at 0.3 photon s$^{-1}$ px$^{-1}$ at our simulated pointing coordinates.

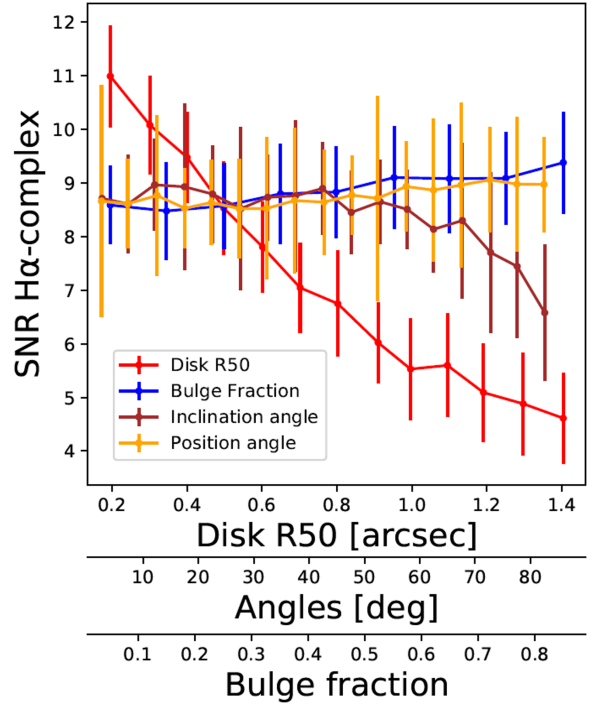

**Fig. 11.** Comparison of the impact of morphological parameters on the S/N of the extracted Hα-complex lines. The results are presented for four sub-samples of 1248 sources with all morphological parameters set at their default value (see Sect. 2.3.2) but varying one parameter at a time. Namely these parameters are the disk R50 (red), the bulge fraction (blue), the inclination angle (brown), and the position angle (orange). The lines and error bars show the median S/N and MAD values calculated in bins of the corresponding morphological parameters.

To measure the emission line fluxes, we fitted a set of Gaussian profiles to the 1D spectrum at the known positions of the lines using the true redshift and left the FWHM and amplitude as free parameters. The continuum is modelled by a linear fit around each line. The S/N for each emission line measurement was defined as the integrated line flux ($S$) over the noise ($N$). The noise was evaluated from the sum in quadrature of the root mean square (RMS) within a $\pm 3\sigma$ region surrounding the central wavelength. To measure the integrated continuum flux, we multiply the extracted spectra with the transmission curve of the $H_E$ filter over the mutual passband covered by the RGS and $H_E$ filter, that is cutting the response of the $H_E$ filter for wavelength above 1.86 μm. The S/N of the continuum measurements was inferred from the ratio of the mean values of the extracted flux and RMS over the wavelength range used to calculate the magnitude. The magnitude derived in this manner is referred in the following by $H$ only since it does not correspond to the full $H_E$ band.

### 5.1. The morphological effects on the quality of the extracted spectra

We present in this section the analysis of the morphological effect on the quality of the slitless spectra referring to the dataset described in Sect. 2.3.2. We only report the results related to the galaxy size, that is the half-light radius of the disk component (disk R50), which has the strongest impact on the quality of the NISP spectra (see Fig. 11). We therefore present our analysis of the extracted spectra obtained from the same source simulated 1248 times with only one variable morphological parameter, the disk R50 that ranges from $0\farcs1$ up to $2''$, all the other parameters being fixed at their default value.

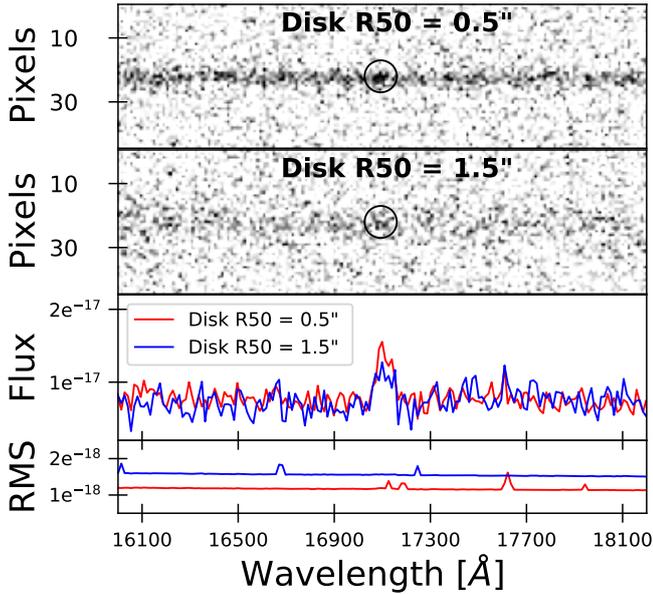

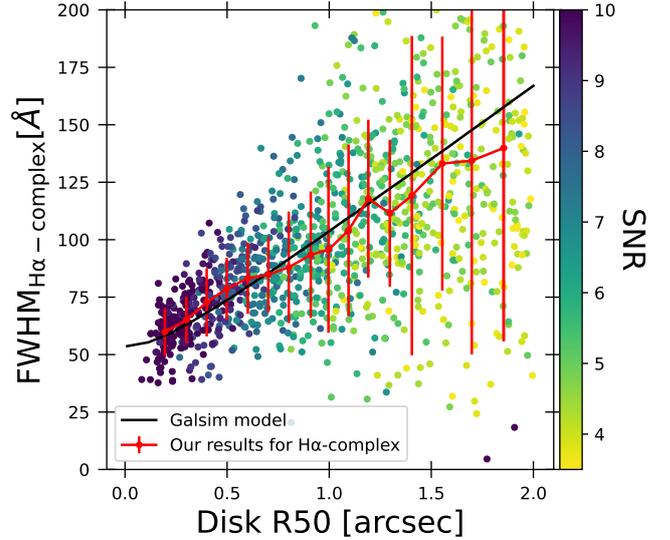

**Fig. 12.** Illustration of the disk R50 effect on the quality of the spectra. *Top two panels*: 2D extracted spectra obtained from one dither of the EWS simulation for two sources with disk R50 = $0\farcs5$ and disk R50 = $1\farcs5$ centred on the H$\alpha$-complex line which is highlighted with a black circle. *Bottom two panels*: the corresponding 1D extracted spectra, resulting from the SIR data processing pipeline on four dithered frames. The signal and RMS spectra are shown and centred on the H$\alpha$-complex line measured with S/N of 9.1 for the $0\farcs5$ source (red line) and 4.5 for the $1\farcs5$ source (blue line). Values of the flux and RMS on the vertical axis are expressed in erg s$^{-1}$ cm$^{-2}$ Å$^{-1}$.

**Fig. 13.** FWHM measured on the extracted H$\alpha$-complex lines as a function of the object disk R50 in arcseconds. The FWHM has been measured on a sample of 1248 sources with all morphological parameters set at their default value (see Sect. 2.3.2) except the disk R50 that ranges from $0\farcs1$ up to $2''$. The red line shows the median FWHM calculated in disk R50 bins including a fixed number of 50 sources. The error bars show the median absolute deviation (MAD). The solid black line shows the model obtained from `GALSIM` (see explanation in the text).

We present in Fig. 12 an overview of the size effect on the extracted spectra by showing the results around the H$\alpha$-complex line obtained emulating twice the same galaxy, that is with same incident spectra but with two different disk R50 set at $0\farcs5$ and $1\farcs5$. In the following sub-sub-sections, we subsequently comment on three effects observable in this figure, namely the increase with the disk R50 of the FWHM in Sect. 5.1.1, the increase in the RMS subsequently decreasing the S/N in Sect. 5.1.2, and the increase in the flux loss related to the extraction procedure and aperture in pixels in Sect. 5.1.3.

### 5.1.1. The spectral resolution of the extracted spectra as a function of the disk R50

To quantify the effect of the size on the spectral resolution, we present in Fig. 13 the measured FWHM of the H$\alpha$-complex[11] as a function of the disk R50. The red line and circles indicate the median FWHM calculated in disk R50 bins including a fixed number of 50 sources. The error bars show the MAD.

As expected, the FWHM scales as the disk R50. We can also see that our measurements are in agreement with the expectation from the model computed with `GALSIM` (Rowe et al. 2015). The model was constructed based on a disk profile matching the properties of the galaxy simulated by TIPS with inclination 45° and oriented at 45° with respect to the dispersion direction. We modelled the H$\alpha$-complex in the two-dimensional spectrum by rendering the galaxy image of each emission line at their respective wavelengths. The line profiles were computed with `GALSIM` considering the velocity dispersion, the shape of the galaxy, that is disk size, inclination and position angles, and convolving with the PSF of the instrument. We then projected the image to make the one-dimensional spectrum and measured the FWHM of the complex. The agreement between the measurement and the model demonstrates that the pipeline successfully recovers the line profile.

---

[11] H$\alpha$ and [N II]$\lambda\lambda6584,6549$ are blended in the extracted *Euclid* RGS spectra.

### 5.1.2. The signal to noise ratio as a function of the disk R50

To evaluate the effect of the size on the quality of the measurement, we characterised the decline of the S/N as the disk R50 increases. Results are shown in Fig. 14 for the H$\alpha$-complex emission line flux measurement (in red) and for the continuum flux measurement in the $H$ band (in green) that we normalised by the median S/N of a $0\farcs25$ disk R50 sources. The red line shows the median normalised S/N calculated in disk R50 bins including a fixed number of 50 sources. The error bars show the MAD.

For both measurements, the curves follow an almost linear drop of the S/N as the size increases, with a smoothing of the S/N drop for disk R50 > $1''$ due to the increase in the extraction aperture in pixels that mitigates the S/N drop. The aperture in pixels is discussed in Sect. 5.1.3. The degradation of S/N reaches ~20% for emission line measurements and ~12% for continuum measurements at the median disk R50 of our EWS and EDS simulation sub-samples, that is disk R50 = $0\farcs4$ and reaches 45% on emission line measurements for $1''$ disk R50 sources.

### 5.1.3. The flux loss due to the extraction procedure and aperture in pixels as a function of the disk R50

The extraction of the 1D spectra in slitless spectroscopy is a challenging process that requires knowledge of the shape of the object as well as of the PSF of the instrument. The SIR pipeline fixes the extraction aperture according to the disk R50 (as shown in the Fig. 15). It is worth noting that the determination of the aperture also takes into account the inclination and position angles, which become increasingly significant as the disk R50 value increases.

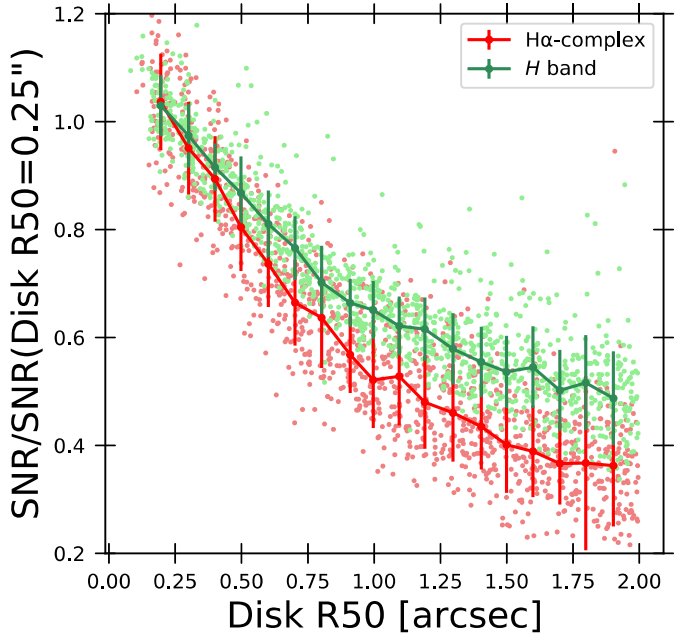

**Fig. 14.** S/N of the extracted Hα-complex measurements (red) and extracted continuum measurements (green) normalised by the median S/N for sources with a disk half-light radius (R50) of 0′′25 as a function of the disk R50. The lines and error bars show the normalised median S/N and MAD values calculated in disk R50 bins including a fixed number of 50 sources. Results are presented for a sub-sample of 1248 sources with all morphological parameters set at their default value (see Sect. 2.3.2) except the disk R50, which varies from 0′′1 to 2′′.

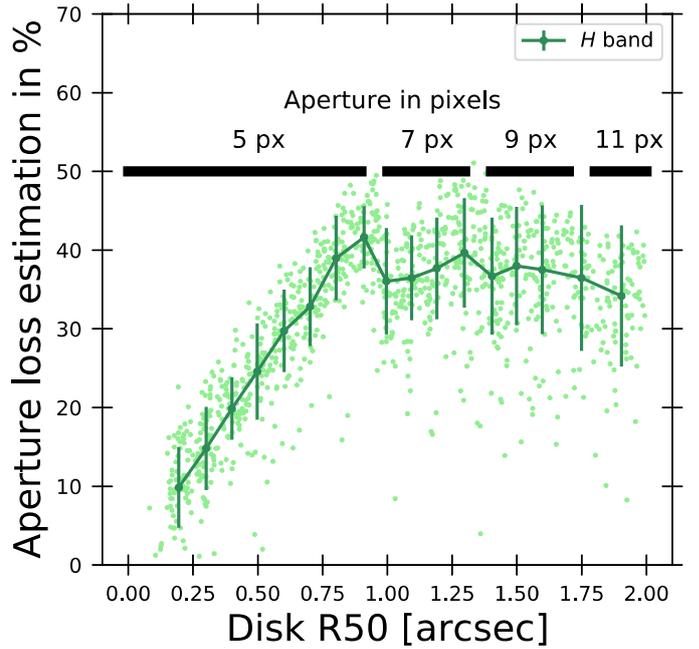

**Fig. 15.** Estimation of the aperture loss as a function of the disk R50 in arcseconds. The line and error bars are obtained calculated the median aperture loss obtained on disk R50 bins including a fixed number of 50 sources. The method to estimated the loss due to the aperture is described in the text. The results are presented for a sub-sample of 1248 sources with all morphological parameters set at their default value (see Sect. 2.3.2) except the disk R50, which varies from 0′′1 up to 2′′. We indicated with black thick lines the range of sizes corresponding to the different apertures in pixels, recalling that the spatial resolution of the NISP instrument is 0′′3.

The estimation of the loss due to the aperture is estimated from the differences between incident and extracted flux obtained convolving the spectra with the $H_E$ filter transmission curve on the mutual wavelength range between the RGS passband and $H_E$. The line indicated in Fig. 15 is the median flux loss calculated in disk R50 bins including a fixed number of 50 sources. The error bars show the MAD. We can see in Fig. 15 that for sources with disk R50 < 0′′25 there is a systematic loss of ~ 10% for the continuum that can be attributed to the loss due to the extraction aperture. To verify this last statement, we performed a simulation increasing the minimal aperture from the default value of five pixels[12] to nine pixels, which brought the loss to less than 1% for the sources with disk R50 < 0′′25. We can also see that the loss due to the aperture increases linearly with the size to reach a loss of ~ 40% for disk R50 ~ 1′′. In this range of disk R50, the aperture set by the pipeline remains at five pixels. For the sources with disk R50 > 1′′, the aperture in pixels increases as indicated in Fig. 15 and a plateau is reached at a loss of ~40% attesting to the fact that for such extended sources an even larger aperture is required to mitigate the flux loss.

The data reduction pipeline for the spectra extraction is still undergoing continuous optimisations to improve the S/N and flux determination. An optimal extraction method will soon be introduced in the SIR pipeline to improve the S/N by assigning weights to pixels based on the fraction of object flux they contain. The optimal extraction will also address and compensate for the aperture effects discussed in this section by applying a 'normalisation factor' that accounts for the flux falling outside the aperture through a cross-dispersion profile inferred for each galaxy.

We used these results to correct the incident versus extracted fluxes of the EWS and EDS simulations for the continuum (see contours Fig. 16) and for the emission lines (see contours Figs. 18 and 19).

### 5.2. The analysis of the extracted spectra of the EWS and EDS simulations

#### 5.2.1. Continuum extraction from the EWS and EDS simulations

We present a comparison of the incident versus extracted magnitudes in the $H$ band in Fig. 16 for the EWS and EDS simulations after applying the correction due to the aperture estimated in Sect. 5.1.3. The colour-coded circles are measurements obtained at $S/N \geq 3.5$ while the grey circles are measurements below that threshold. The contours are the iso-proportions of the distribution density of our measurements before applying any correction for the loss due to the extraction aperture. We can see from the scattered circles in Fig. 16 that we found an almost perfect match between the incident and extracted measured magnitudes once the fluxes are corrected for the estimated size-dependent flux loss due to the aperture extraction.

---

[12] This minimal aperture in pixels was determined by the OU-SIR team as an optimal trade-of based on S/N analysis.

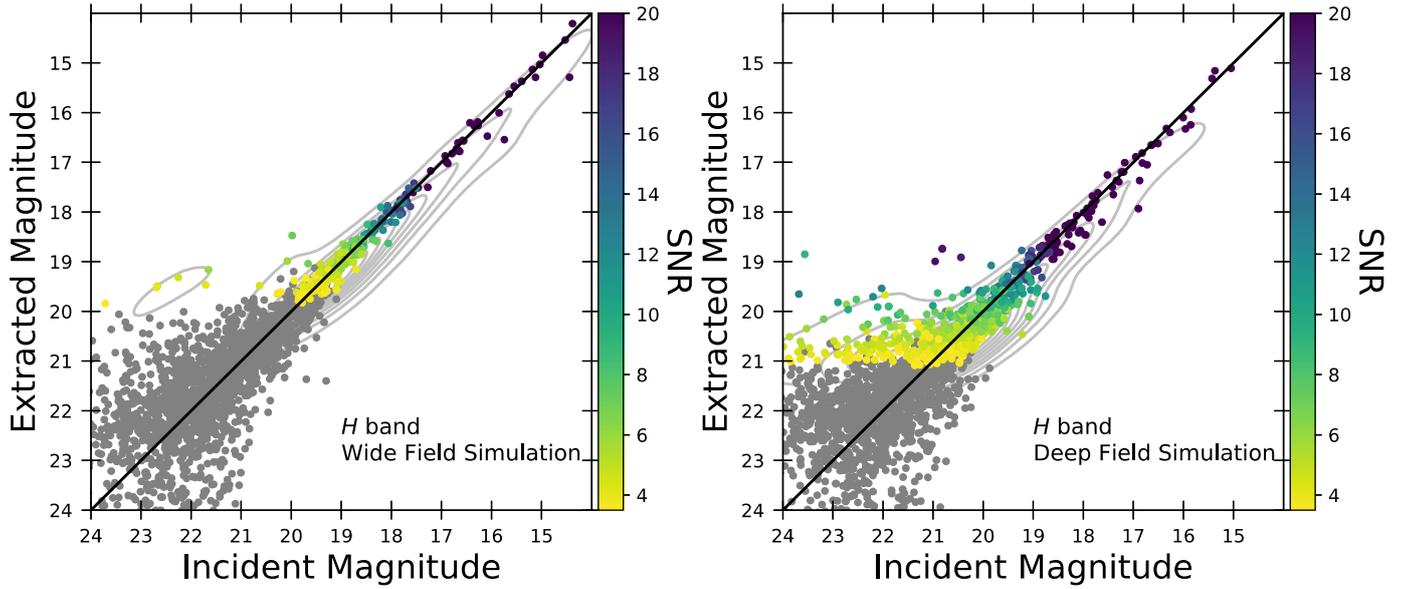

**Fig. 16.** Results of the EWS simulation (*left*) and of the EDS simulation (*right*) comparing the measured extracted versus incident $H$ band magnitudes calculated in the AB system. The magnitude is obtained convolving the extracted and incident spectra with the transmission curve of the $H_E$ filter on the mutual wavelength range between the RGS passband and $H_E$. The scattered circles are corrected for the estimated loss due to the aperture (see Sect. 5.1.3). The sources are colour-coded with the S/N, as shown by the right-side colour bar, and the grey circles are objects with $S/N < 3.5$. The black diagonal line shows the 1:1 ratio. The contours are the iso-proportions of the distribution density of our measurement, starting at 20% with a 10% step, before applying the correction due to the aperture loss.

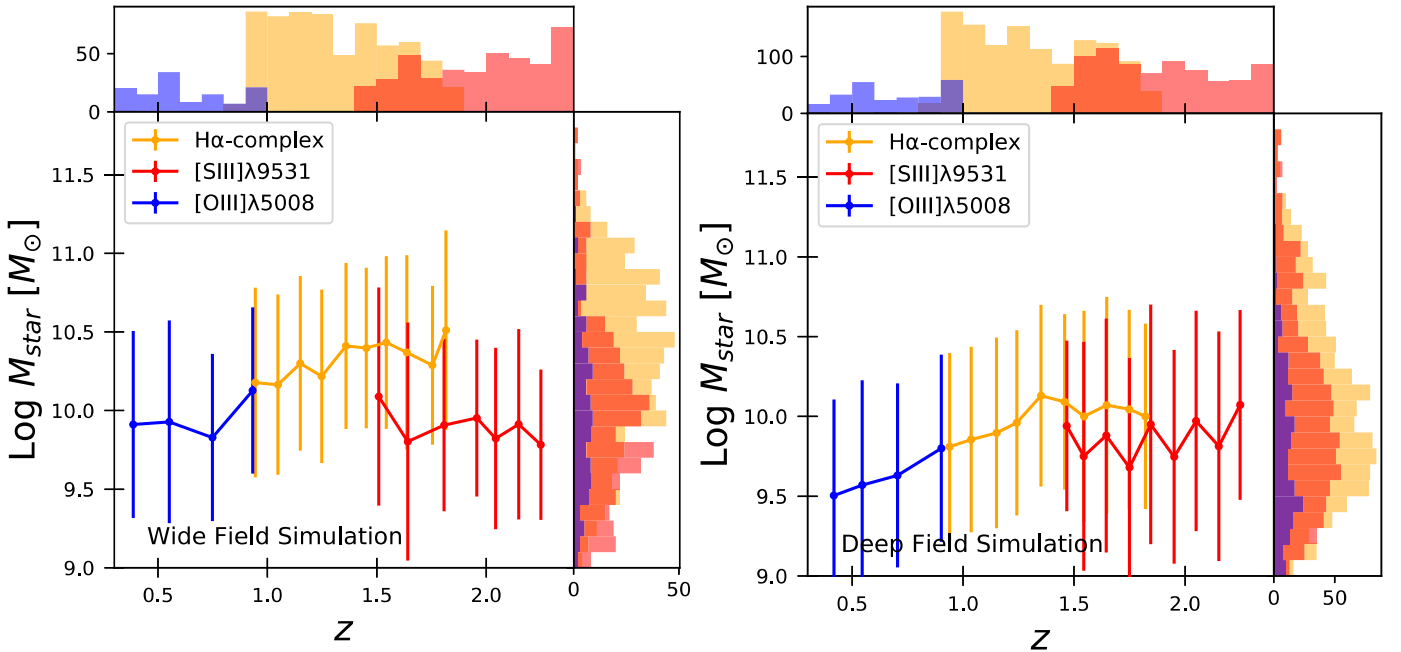

**Fig. 17.** Distribution of the sources with an emission line measurement at $S/N \geq 3.5$, in the stellar mass versus redshift plane. Results are indicated for the EWS (*left*) and for the EDS (*right*) simulations and for the H$\alpha$-complex (orange), [O III]$\lambda5008$ (blue) and [S III]$\lambda9531$ (red) emission lines. The lines and error bars show the median S/N and MAD values calculated in redshift bins including a fixed number of 50 sources.

### 5.2.2. Emission lines extraction from the EWS and EDS simulations

We present in Fig. 17 the distribution of the sources with an emission line measurement at $S/N \geq 3.5$ in the stellar mass versus redshift plane. We indicate the distribution corresponding to measurements of the [S III]$\lambda9531$, H$\alpha$, and [O III]$\lambda5008$ emission lines. In the redshift range where both H$\alpha$ and [O III]$\lambda5008$ emission lines fall in the RGS passband, that is at $1.5 \leq z \leq 1.8$, we can see that the median stellar mass of H$\alpha$ emitters is higher than the median stellar mass of [O III]$\lambda5008$ emitters. This effect is related to the anti-correlation between the mass and [OIII]$\lambda5008$ emission line flux due to a relatively lower ionising power for galaxies with higher metallicities (see Colbert et al. 2013; Domínguez et al. 2013, for further explanation on this observed spectral feature). On the contrary, the H$\alpha$ emission line is expected to correlate with the mass due to the SFR-H$\alpha$ relation (see Kennicutt 1998, and Sect. 2.4 for the distribution of SFGs on the $M_\star$-SFR plane). This aspect is particularly relevant to make forecast on number counts of galaxies with two detected emission lines. On the other hand and of interest for galaxy evolutionary studies, a related effect on the power of the stacking analysis is presented in Sect. 5.4.

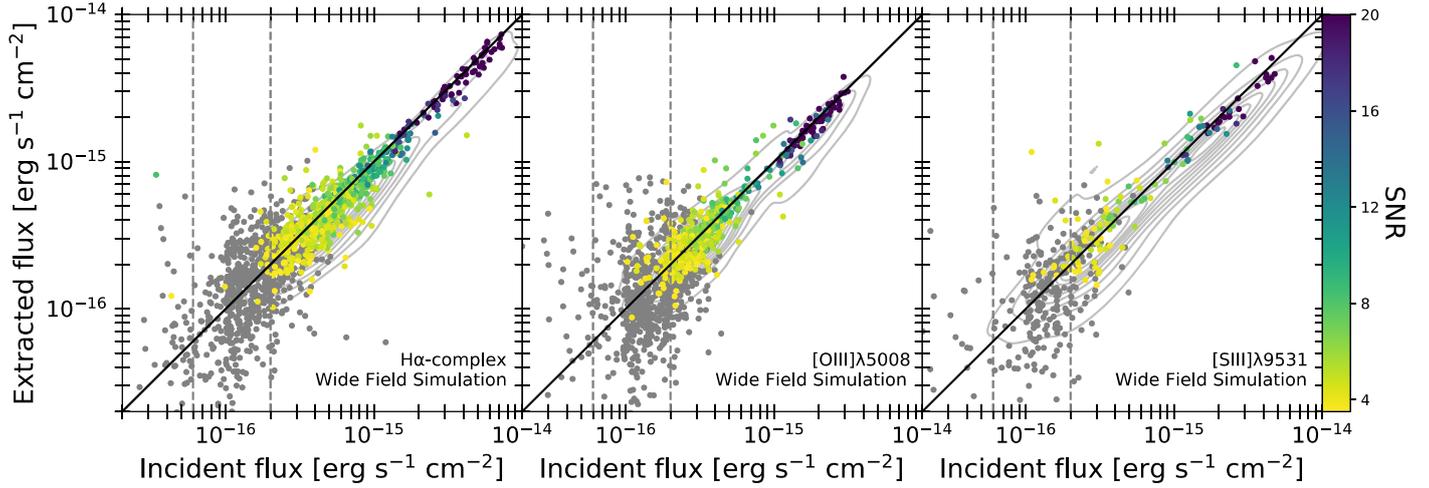

**Fig. 18.** Results of the EWS simulation comparing the extracted versus incident fluxes of the H$\alpha$-complex (*left*), [O III]$\lambda$5008 (*middle*) and [S III]$\lambda$9531 (*right*) emission lines. The scattered circles are corrected for the estimated loss due to the aperture (see Sect. 5.1.3). The sources are colour-coded with the S/N of the emission line measurement, as shown by the right-side colour bar. The grey circles are objects with $S/N <$ 3.5. The black diagonal line shows the 1:1 ratio. The grey dotted lines are the requirements for the *Euclid* spectroscopic detection limit for the EDS (leftmost) and EWS (rightmost) surveys. The contours correspond to iso-proportions of the distribution density of our measurements, starting at 20% with a 10% step, before applying the correction due to the aperture loss.

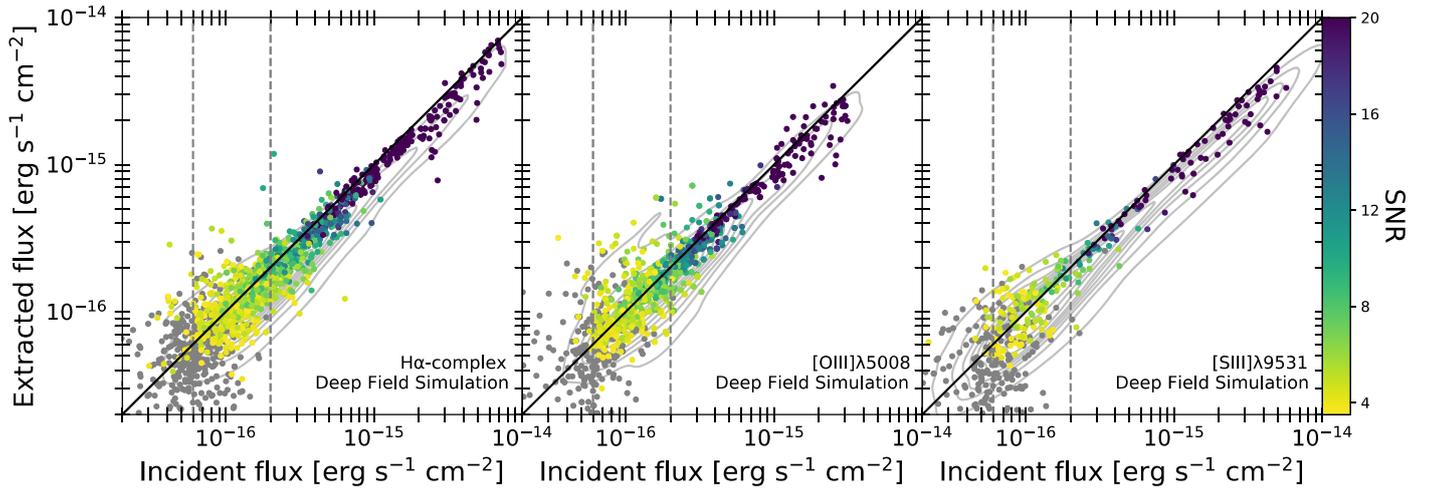

**Fig. 19.** Same plots as in Fig. 18, but for the EDS simulation.

We present in Fig. 18 for the EWS simulation and in Fig. 19 for the EDS simulation, a comparison between the incident and extracted fluxes for the H$\alpha$-complex (left panel), [O III]$\lambda$5008 (middle panel) and [S III]$\lambda$9531 (right panel) emission lines. The scattered circles are corrected for the loss due to the extraction aperture estimated in Sect. 5.1.3. The colour-coded circles are the measurements at $S/N \geq 3.5$ while the grey circles indicate objects with S/N below that threshold. The *Euclid* requirements for emission line detection limits at $S/N = 3.5$ for $0\farcs25$ radius extended sources, set as $2 \times 10^{-16}$ erg s$^{-1}$ cm$^{-2}$ for the EWS (Scaramella et al. 2022), and at $6 \times 10^{-17}$ erg s$^{-1}$ cm$^{-2}$ for EDS, are indicated with the dashed grey lines on the figures. The contours are the iso-proportions of the distribution density of our measurements before applying the correction due to the aperture loss. As for the continuum measurements, we found a promising agreement between the incident and extracted fluxes.

In the EWS, we found that 35 out of the 600 H$\alpha$ emission lines measured with $S/N \geq 3.5$ have also the continuum detected with $S/N \geq 3.5$. Similarly, in the EDS, 299 out of the 1161 H$\alpha$ emission lines measured with $S/N \geq 3.5$ have the continuum detected.

### 5.3. Estimation of the NISP-S detection limit

We present in Fig. 20 (left panel) the S/N measured on the extracted continuum versus the incident magnitude in the $H$ band. The results of the EWS simulation are indicated in red and the results of the EDS simulation are indicated in blue. The corresponding lines are the median S/N calculated in bins of magnitudes including a fixed number of 50 sources. The error bars show the MAD. The dashed black line indicates the $S/N$ of 3.5. We found a continuum detection limit at $S/N = 3.5$ for the EWS simulation at $H = 19.5 \pm 0.2$ mag and for the EDS simulation at $H = 20.8 \pm 0.6$ mag. These limits are found for our sub-samples that contain galaxies with a median disk R50 of $0\farcs4$. The difference between the EWS and EDS inferred detection limit agrees with the expected 1.25 mag increasing depth expected for the EDS, which corresponds to an exposure time ten times longer.

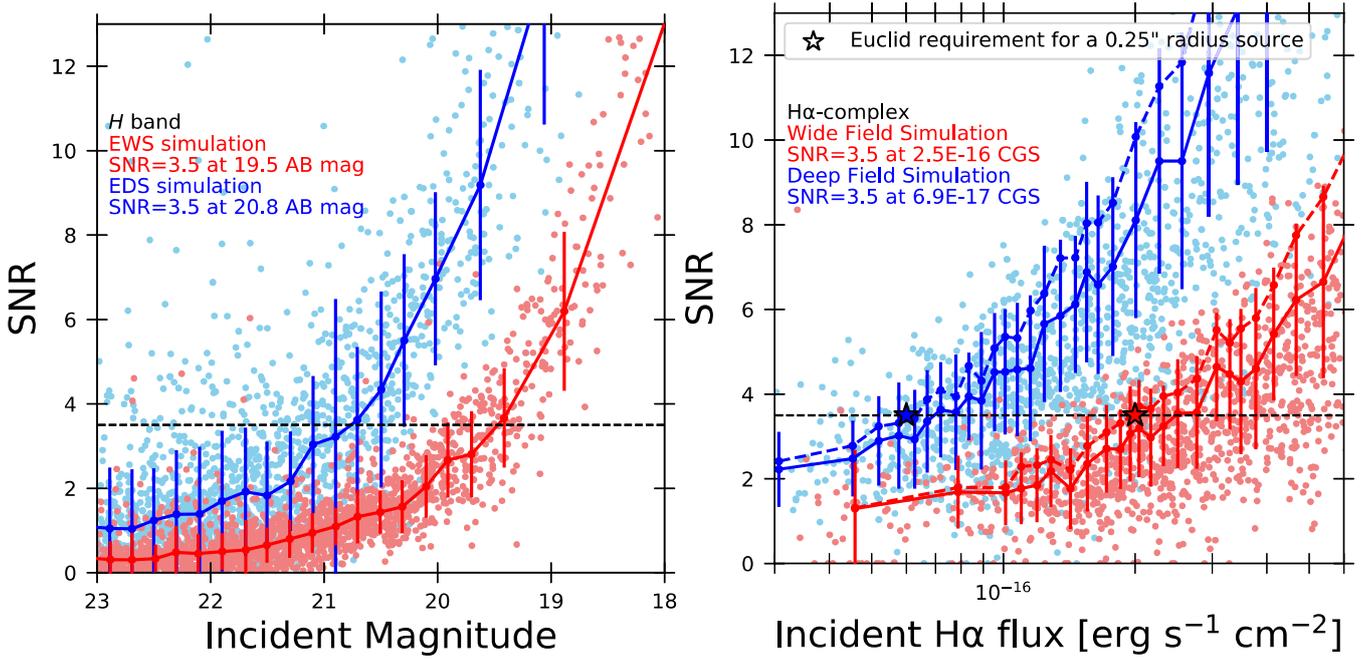

**Fig. 20.** Results of the NISP detection capabilities for the continuum and emission lines. *Left*: S/N of the continuum measurements versus the incident magnitude in the $H$ band. *Right*: S/N of the H$\alpha$-complex measurements versus the incident H$\alpha$ flux. Results are indicated for the EWS simulation in red and for the and EDS simulation in blue. The red and blues lines are the median S/N calculated in incident magnitude (*left*) and H$\alpha$ flux (*right*) bins including a fixed number of 50 sources. The error bars show the MAD. The dashed red and blue lines in the *right* panel indicate the median value corrected for the size effect on the S/N (see Sect. 5.1.2). The dashed black line indicates the S/N of 3.5 which we used to infer the detection limits for the EWS and EDS indicated on the figure and reported in Table 2.

We present in Fig. 20 (right panel) the S/N of the measurement on the H$\alpha$-complex line as a function of the incident H$\alpha$ flux. The results for the EWS simulation are indicated in red and the results for the EDS simulation are indicated in blue. The corresponding lines are the median S/N calculated in bins of incident H$\alpha$ flux including a fixed number of 50 sources. The error bars show the MAD. We indicated with the dashed black line the S/N of 3.5 and with star markers the *Euclid* RGS requirement for a $0\farcs25$ radius source. We found an emission line detection limit at $S/N = 3.5$ at $(2.5 \pm 0.6) \times 10^{-16}$ erg s$^{-1}$ cm$^{-2}$ for the EWS and at $(6.9 \pm 2.8) \times 10^{-17}$ erg s$^{-1}$ cm$^{-2}$ for the EDS. The difference in flux sensitivity between our EWS and EDS detection limits is in agreement with the expected increase that scales as the square root of the amount of dithered frames, that is $\sqrt{10}$. We recall that these limits are found for our sub-samples that contain galaxies with a median disk R50 of $0\farcs4$. For comparison with the *Euclid* RGS H$\alpha$ detection limit requirement made for a $0\farcs25$ radius source (indicated with star markers), we corrected the results for the disk R50 effect on the S/N to provide predictions for a source with a disk R50 of $0\farcs25$ (see Sect. 5.1.2) and found a very good agreement with the *Euclid* requirement. The corrected results are indicated with the dashed red and blue lines on the figure. We present in Table 2 a summary of the continuum and emission lines detection limits obtained for the EWS and EDS simulations.

**Table 2.** Results obtained at $S/N = 3.5$ for the measurement on the extracted spectra of the EWS and EDS simulations containing sources with median disk R50 of $0\farcs4$.

|     | Exposure time (s) | Continuum $H$ band (mag) | Emission lines H$\alpha$ (CGS) |
| --- | --- | --- | --- |
| EWS | 2212 | $19.5 \pm 0.2$ | $(2.5 \pm 0.6) \times 10^{-16}$ <br> *Req.: $2 \times 10^{-16}$* |
| EDS | 22 120 | $20.8 \pm 0.6$ | $(6.9 \pm 2.8) \times 10^{-17}$ <br> *Req.: $6 \times 10^{-17}$* |

**Notes.** The results of the continuum detection limit are indicated in AB magnitudes in the $H$ band. The results of the H$\alpha$ detection limit are in erg s$^{-1}$ cm$^{-2}$. The *Euclid* requirement for a $0\farcs25$ disk R50 source is indicated in italic.

### 5.4. Stacking analysis on the extracted spectra

We present in this section a stacking analysis that has been performed co-adding extracted spectra of different sources taken from our EWS simulation dataset. We assumed that redshifts are well determined from the H$\alpha$-complex emission line, by only considering objects with S/N(H$\alpha$-complex) $\geq 3.5$. We also assumed that the stellar mass will be well constrained from the SED fitting analysis on photometric data from *Euclid* and external ground-based instruments (e.g. Rubin/LSST, DECAM, Subaru; see Scaramella et al. 2022).

In this analysis we highlight the effect of the stellar mass on the stacking analysis performance. To do so, we performed a stacking analysis on two $M_\star$ bins. The first bin contains sources with $9.5 \leq \log_{10}(M_\star [M_\odot]) \leq 10.0$ (hereafter referred as sample 1). The second bin contains galaxies with $10.5 \leq \log_{10}(M_\star\,[M_\odot]) \leq 11.0$ (hereafter referred as sample 2). We selected galaxies located at $1.5 \leq z \leq 1.8$, that is the redshift range where H$\alpha$ and [O III]$\lambda$5008 lines fall in the RGS passband, with S/N(H$\alpha$-complex) $\geq 3.5$ and having [O III]$\lambda$5008 measurements with $S/N < 3.5$.

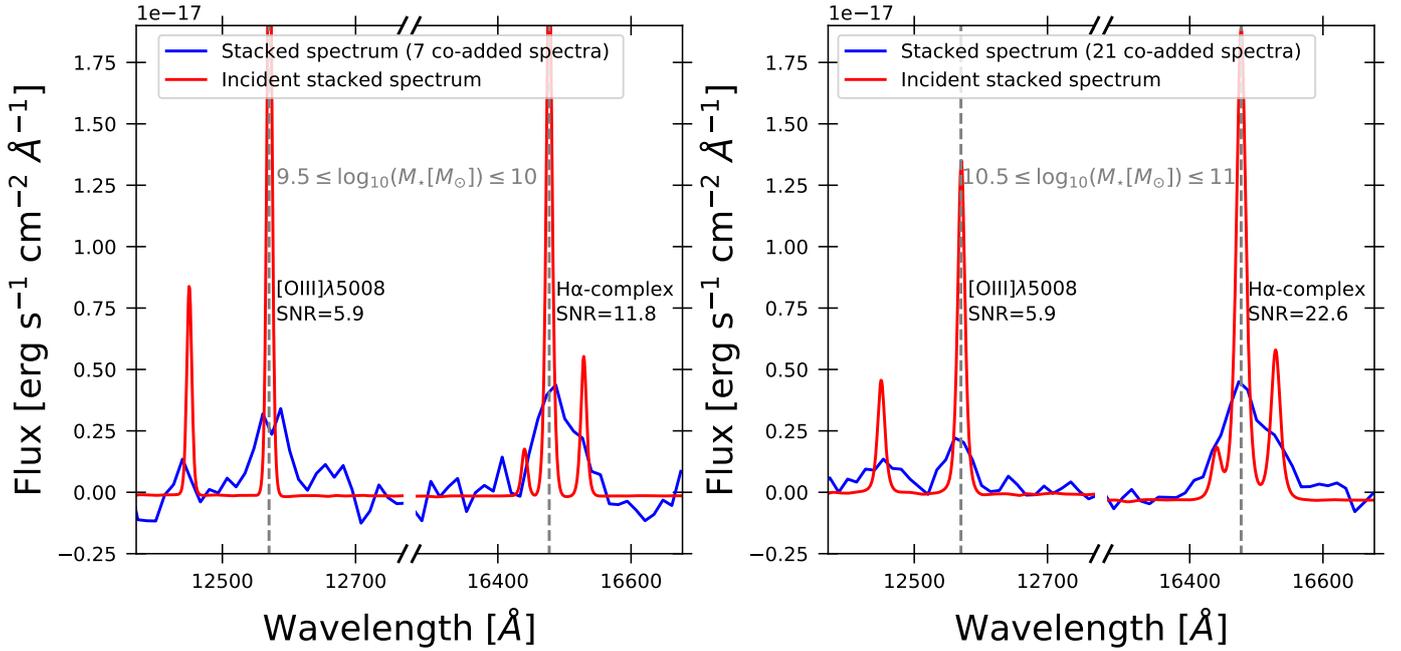

**Fig. 21.** Results of the stacking analysis performed on two H$\alpha$-detected samples with different total stellar mass ($M_\star$) unveiling the [O III]$\lambda$5008 otherwise too faint to be detected on the single co-added spectra. *Left*: stacking analysis on seven co-added spectra with $9.5 \leq \log_{10}(M_\star\,[M_\odot]) \leq 10.0$. *Right*: stacking analysis on 21 co-added spectra with $10.5 \leq \log_{10}(M_\star\,[M_\odot]) \leq 11.0$.

Following the stacking method for sources located at different redshift described by Spilker et al. (2014), each co-added spectrum has been shifted in redshift to a common redshift that we chose to be the lowest redshift ($z_{\min}$) of the samples 1 and 2, at $z_{\min} = 1.51$ and $z_{\min} = 1.52$ respectively. Each spectrum is resampled on a common wavelength array with the linear dispersion of the source located at $z_{\min}$, corresponding to the worst spectral sampling among the co-added stacked objects. The flux and wavelengths of the extracted spectra are scaled accordingly, with a similar procedure as described in Sect. 3.1. We then performed the stacking analysis calculating the weighted mean of the fluxes assuming a Gaussian behaviour for the noise. The RMS spectrum obtained co-adding the spectra scales as the inverse of the square root of the number of co-added spectra.

We present in Fig. 21 the results of the stacking analysis performed on the sample 1 and sample 2. We found that sample 2 requires about three times more co-added spectra than the sample 1 to obtain a similar S/N([O III]$\lambda$5008). The factor three comes from a combination of two effects. First, by construction, the [O III]$\lambda$5008/H$\alpha$ of our simulated galaxies decreases with increasing stellar mass (and H$\alpha$ luminosity), in agreement with what observed for SFGs at similar redshifts (Domínguez et al. 2013; Colbert et al. 2013). Figure 17 illustrates the median stellar mass difference between the H$\alpha$ and [O III]$\lambda$5008 bright emitters, that is with the emission line measured at $S/N \geq 3.5$. This makes it more difficult to detect the [O III]$\lambda$5008 emission from galaxies in the most massive sample 2. Second, as a consequence of the $M_\star$-size relation, the average half-light radius of sample 2 (disk R50 $\sim$ 0″.6) is about two times larger than that of sample 1 (disk R50 $\sim$ 0″.3). This also contributes to an increase in the number of co-added spectra needed to detect the [O III]$\lambda$5008 emission line in the stacking analysis due to the drop in S/N with increasing galaxy size, as shown in Sect. 5.1.2. To quantify the effect of the size, we also performed the stacking on sources with similar H$\alpha$ and [O III]$\lambda$5008 fluxes, split in two sub-samples with different average sizes (disk R50 $\leq$ 0″.3 versus disk R50 $\geq$ 0″.5). In this case, we found that two times the number of galaxies in the large-size bin are needed compared to galaxies in the smaller size bin in order to obtain similar S/N for the [O III]$\lambda$5008 emission line. This exercise reminds us that size effects should be fully taken into account, for example in the sample selection, in order to optimise the results and to avoid possible biases when stacking slitless spectra. Beyond these caveats, this analysis also shows the potential of the stacking approach to characterise the average interstellar medium (ISM) properties of galaxies through emission lines diagnostics even when only one line, for example H$\alpha$, is above the detection limit of the RGS.

## 6. Caveats

This work is a preliminary study of the NISP-S capabilities and some caveats related to the simulation setup are to be kept in mind when interpreting the results presented in this paper. We summarise these caveats as follows.

We avoided the contamination due to the overlap of the spectra and simulated only the first order spectra. Previous analysis have shown that the decontamination of slitless spectra can be well addressed when a direct image provides information on the positions, sizes and spectral shapes, and particularly when dispersed images with different grism orientations are available (Walsh et al. 2010; Momcheva et al. 2016; Ryan et al. 2018), as will be the case for *Euclid*.

We assumed a constant spectral dispersion on the field of view even if during the ground-test campaigns, using a Fabry-Perot etalon (see Gabarra 2023; Gillard et al., in prep.), the spectral dispersion varied according to the position on the detector. Our choice is justified by the additional phenomena that will alter the in-orbit spectral dispersion, for example launch vibrations, zero gravity conditions, and thermal stabilisation. The in-flight spectral dispersion calibration will be performed using compact planetary nebulae with known emission lines (Paterson et al., in prep.).

We considered the same size and shape in both the continuum and emission lines. This choice is motivated by the strong correlation shown by Bagley et al. (2020) between the size of galaxies in the continuum and H$\alpha$ emission line in the redshift range targeted by *Euclid*.

## 7. Conclusions

In this work, we have presented the construction and simulation of spectroscopic data emulating those that will be produced by the RGS of the *Euclid* NISP spectrograph. We started by presenting the publicly released catalogues from the COSMOS and CANDELS/GOODS-N fields. We then built spectral energy distributions based on stellar population models using the software `GALAXEV`, referring to the physical properties available for each observed galaxy. Separately, we calculated the nebular emission line fluxes using observed scaling relations and photoionisation models as described in Sect. 3. Other efforts to predict emission lines using an alternative method based on cosmological simulations and photoionization models is currently ongoing (Hirschmann et al, in prep). A cross-check between these two approaches will be of great use in the context of the *Euclid* efforts for galaxy evolution studies.

We created two datasets aimed at providing realistic sub-samples of the sources that will be detected by the RGS spectrograph in the EWS and EDS. The median disk R50 in the EWS and EDS sub-samples is $0\farcs4$. To probe the effects of galaxy morphology expected to be important in the slitless spectroscopy, we also built a dedicated set of simulations to characterise the impact of the angular size on the extracted spectra.

This work represents the first step in a campaign test to assess the performance of the NISP spectrograph in the context of galaxy evolution studies and the main results from the analysis can be summarised as follows:

1. The morphological parameter that most affects the quality of the extracted spectrum from the NISP instrument is the half-light radius. The increase in the half-light radius has the effect to degrade the spectral resolution, to decrease the S/N, and to increase the flux loss due to the extraction aperture. In particular, we found that the emission line S/N drops by ~45% when the disk R50 ranges from $0\farcs25$ to $1''$.
2. With the simulated background levels described in Sect. 4, we inferred the 3.5$\sigma$ NISP spectroscopic detection limit for the continuum at $H = 19.5 \pm 0.2$ mag for the EWS simulation and at $H = 20.8 \pm 0.6$ mag for the EDS simulation.
3. We have presented the median stellar mass of the sources expected to be detected as a function of redshift and of the targeted emission lines. For example, at $z = 1.6$ in the EWS configuration, we show that *Euclid* will detect the H$\alpha$ emission line emitted for sources with a median $\log_{10}(M_\star) \sim 10.5$ while the [O III]$\lambda$5008 emission line will be detected for sources with a median $\log_{10}(M_\star) \sim 10.0$. This effect, which is related to the MZR, is of crucial importance for future *Euclid* science, for example to make a forecast as to the number counts of galaxies with two detected emission lines.
4. We found the 3.5$\sigma$ detection limit for emission lines to be $(2.5 \pm 0.6) \times 10^{-16}$ erg s$^{-1}$ cm$^{-2}$ for the EWS and $(6.9 \pm 2.8) \times 10^{-17}$ erg s$^{-1}$ cm$^{-2}$ for the EDS. We corrected these results for the size effect and found a very good agreement with the *Euclid* RGS emission line detection limit requirement, which are expressed for a $0\farcs25$ radius source.
5. We characterised the effect of the stellar mass on the potential for the stacking analysis to unveil the [O III]$\lambda$5008 emission line otherwise too faint to detect and we show that about three times more co-added EWS extracted spectra are required for galaxies with $10.5 \leq \log_{10}(M_\star [M_\odot]) \leq 11.0$ compared to galaxies with $9.5 \leq \log_{10}(M_\star [M_\odot]) \leq 10.0$. This factor of three is the result of the fact that [O III]$\lambda$5008/H$\alpha$ is lower and that the disk R50 is larger for the more massive galaxies. This analysis also shows that stacking *Euclid* spectra will be very promising to characterise the average ISM properties of galaxies when only one line, for example H$\alpha$, is above the detection limit.

This pilot run represents an effort to assess the *Euclid* spectroscopic capabilities to detect SFGs and a coming paper (Lusso et al., in prep.) will present a similar study for AGNs. More so, upcoming simulations are being planned with a larger sample of galaxies, sampling different regions of the sky, and including the blue grism observations for the EDS.

*Acknowledgements.* We thank Paolo Cassata, Alberto Franceschini, Salvatore Quai, Alvio Renzini, and Gregory Setnikar for useful discussion. We also thank the anonymous referee for the constructive comments that helped improve this work. The *Euclid* Consortium acknowledges the European Space Agency and a number of agencies and institutes that have supported the development of *Euclid*, in particular the Academy of Finland, the Agenzia Spaziale Italiana, the Belgian Science Policy, the Canadian *Euclid* Consortium, the French Centre National d'Etudes Spatiales, the Deutsches Zentrum für Luft- und Raumfahrt, the Danish Space Research Institute, the Fundação para a Ciência e a Tecnologia, the Ministerio de Ciencia e Innovación, the National Aeronautics and Space Administration, the National Astronomical Observatory of Japan, the Netherlandse Onderzoekschool Voor Astronomie, the Norwegian Space Agency, the Romanian Space Agency, the State Secretariat for Education, Research and Innovation (SERI) at the Swiss Space Office (SSO), and the United Kingdom Space Agency. A complete and detailed list is available on the *Euclid* web site (http://www.euclid-ec.org).

[1] Dipartimento di Fisica e Astronomia "G. Galilei", Universitá di Padova, Via Marzolo 8, 35131 Padova, Italy
 e-mail: louis.gabarra@pd.infn.it
[2] INFN-Padova, Via Marzolo 8, 35131 Padova, Italy
[3] INAF-IASF Milano, Via Alfonso Corti 12, 20133 Milano, Italy
[4] INAF-Osservatorio Astronomico di Padova, Via dell'Osservatorio 5, 35122 Padova, Italy
[5] Dipartimento di Fisica e Astronomia "Augusto Righi" – Alma Mater Studiorum Università di Bologna, via Piero Gobetti 93/2, 40129 Bologna, Italy
[6] INAF – Osservatorio di Astrofisica e Scienza dello Spazio di Bologna, Via Piero Gobetti 93/3, 40129 Bologna, Italy
[7] Aix-Marseille Université, CNRS/IN2P3, CPPM, Marseille, France
[8] INAF-Osservatorio Astronomico di Brera, Via Brera 28, 20122 Milano, Italy
[9] Dipartimento di Fisica, Università di Genova, Via Dodecaneso 33, 16146 Genova, Italy
[10] INFN-Sezione di Genova, Via Dodecaneso 33, 16146 Genova, Italy
[11] Dipartimento di Fisica e Astronomia, Universitá di Bologna, Via Gobetti 93/2, 40129 Bologna, Italy
[12] INAF-Osservatorio Astronomico di Capodimonte, Via Moiariello 16, 80131 Napoli, Italy
[13] Université Paris Cité, CNRS, Astroparticule et Cosmologie, 75013 Paris, France
[14] Institute of Physics, Laboratory for Galaxy Evolution, Ecole Polytechnique Fédérale de Lausanne, Observatoire de Sauverny, 1290 Versoix, Switzerland
[15] INAF – Osservatorio Astronomico di Trieste, Via G. B. Tiepolo 11, 34143 Trieste, Italy
[16] Department of Mathematics and Physics, Roma Tre University, Via della Vasca Navale 84, 00146 Rome, Italy
[17] Institut d'Astrophysique de Paris, UMR 7095, CNRS, and Sorbonne Université, 98 bis boulevard Arago, 75014 Paris, France
[18] Max-Planck-Institut für Astronomie, Königstuhl 17, 69117 Heidelberg, Germany

19. NASA Goddard Space Flight Center, Greenbelt, MD 20771, USA
20. Université Paris-Saclay, CNRS, Institut d'astrophysique spatiale, 91405 Orsay, France
21. Institute of Cosmology and Gravitation, University of Portsmouth, Portsmouth PO1 3FX, UK
22. INFN-Sezione di Bologna, Viale Berti Pichat 6/2, 40127 Bologna, Italy
23. Max Planck Institute for Extraterrestrial Physics, Giessenbachstr. 1, 85748 Garching, Germany
24. Universitäts-Sternwarte München, Fakultät für Physik, Ludwig-Maximilians-Universität München, Scheinerstrasse 1, 81679 München, Germany
25. INAF-Osservatorio Astrofisico di Torino, Via Osservatorio 20, 10025 Pino Torinese (TO), Italy
26. INFN-Sezione di Roma Tre, Via della Vasca Navale 84, 00146 Roma, Italy
27. Department of Physics "E. Pancini", University Federico II, Via Cinthia 6, 80126 Napoli, Italy
28. Instituto de Astrofísica e Ciências do Espaço, Universidade do Porto, CAUP, Rua das Estrelas, 4150-762 Porto, Portugal
29. Dipartimento di Fisica, Universitá degli Studi di Torino, Via P. Giuria 1, 10125 Torino, Italy
30. INFN-Sezione di Torino, Via P. Giuria 1, 10125 Torino, Italy
31. Institut de Física d'Altes Energies (IFAE), The Barcelona Institute of Science and Technology, Campus UAB, 08193 Bellaterra (Barcelona), Spain
32. Port d'Informació Científica, Campus UAB, C. Albareda s/n, 08193 Bellaterra (Barcelona), Spain
33. Institut d'Estudis Espacials de Catalunya (IEEC), Carrer Gran Capitá 2–4, 08034 Barcelona, Spain
34. Institute of Space Sciences (ICE, CSIC), Campus UAB, Carrer de Can Magrans, s/n, 08193 Barcelona, Spain
35. INAF-Osservatorio Astronomico di Roma, Via Frascati 33, 00078 Monteporzio Catone, Italy
36. INFN section of Naples, Via Cinthia 6, 80126 Napoli, Italy
37. Centre National d'Etudes Spatiales – Centre spatial de Toulouse, 18 avenue Edouard Belin, 31401 Toulouse Cedex 9, France
38. Institut national de physique nucléaire et de physique des particules, 3 rue Michel-Ange, 75794 Paris Cedex 16, France
39. Institute for Astronomy, University of Edinburgh, Royal Observatory, Blackford Hill, Edinburgh EH9 3HJ, UK
40. Jodrell Bank Centre for Astrophysics, Department of Physics and Astronomy, University of Manchester, Oxford Road, Manchester M13 9PL, UK
41. ESAC/ESA, Camino Bajo del Castillo, s/n., Urb. Villafranca del Castillo, 28692 Villanueva de la Cañada, Madrid, Spain
42. European Space Agency/ESRIN, Largo Galileo Galilei 1, 00044 Frascati, Roma, Italy
43. University of Lyon, Univ. Claude Bernard Lyon 1, CNRS/IN2P3, IP2I Lyon, UMR 5822, 69622 Villeurbanne, France
44. Aix-Marseille Université, CNRS, CNES, LAM, Marseille, France
45. Institute of Physics, Laboratory of Astrophysics, Ecole Polytechnique Fédérale de Lausanne (EPFL), Observatoire de Sauverny, 1290 Versoix, Switzerland
46. Departamento de Física, Faculdade de Ciências, Universidade de Lisboa, Edifício C8, Campo Grande, 1749-016 Lisboa, Portugal
47. Instituto de Astrofísica e Ciências do Espaço, Faculdade de Ciências, Universidade de Lisboa, Campo Grande, 1749-016 Lisboa, Portugal
48. Department of Astronomy, University of Geneva, ch. d'Ecogia 16, 1290 Versoix, Switzerland
49. Université Paris-Saclay, Université Paris Cité, CEA, CNRS, Astrophysique, Instrumentation et Modélisation Paris-Saclay, 91191 Gif-sur-Yvette, France
50. Istituto Nazionale di Fisica Nucleare, Sezione di Bologna, Via Irnerio 46, 40126 Bologna, Italy
51. Dipartimento di Fisica "Aldo Pontremoli", Universitá degli Studi di Milano, Via Celoria 16, 20133 Milano, Italy
52. INFN-Sezione di Milano, Via Celoria 16, 20133 Milano, Italy
53. Jet Propulsion Laboratory, California Institute of Technology, 4800 Oak Grove Drive, Pasadena, CA 91109, USA
54. Technical University of Denmark, Elektrovej 327, 2800 Kgs. Lyngby, Denmark
55. Cosmic Dawn Center (DAWN), Denmark
56. Institut d'Astrophysique de Paris, 98bis Boulevard Arago, 75014 Paris, France
57. Mullard Space Science Laboratory, University College London, Holmbury St Mary, Dorking, Surrey RH5 6NT, UK
58. Université de Genève, Département de Physique Théorique and Centre for Astroparticle Physics, 24 quai Ernest-Ansermet, 1211 Genève 4, Switzerland
59. Department of Physics, PO Box 64, 00014 University of Helsinki, Finland
60. Helsinki Institute of Physics, Gustaf Hällströmin katu 2, University of Helsinki, Helsinki, Finland
61. Institute of Theoretical Astrophysics, University of Oslo, PO Box 1029 Blindern, 0315 Oslo, Norway
62. NOVA optical infrared instrumentation group at ASTRON, Oude Hoogeveensedijk 4, 7991PD, Dwingeloo, The Netherlands
63. Argelander-Institut für Astronomie, Universität Bonn, Auf dem Hügel 71, 53121 Bonn, Germany
64. Department of Physics, Institute for Computational Cosmology, Durham University, South Road DH1 3LE, UK
65. Université Côte d'Azur, Observatoire de la Côte d'Azur, CNRS, Laboratoire Lagrange, Bd de l'Observatoire, CS 34229, 06304 Nice Cedex 4, France
66. School of Mathematics and Physics, University of Surrey, Guildford, Surrey GU2 7XH, UK
67. European Space Agency/ESTEC, Keplerlaan 1, 2201 AZ Noordwijk, The Netherlands
68. Department of Physics and Astronomy, University of Aarhus, Ny Munkegade 120, 8000 Aarhus C, Denmark
69. Centre for Astrophysics, University of Waterloo, Waterloo, Ontario N2L 3G1, Canada
70. Department of Physics and Astronomy, University of Waterloo, Waterloo, Ontario N2L 3G1, Canada
71. Perimeter Institute for Theoretical Physics, Waterloo, Ontario N2L 2Y5, Canada
72. Space Science Data Center, Italian Space Agency, via del Politecnico snc, 00133 Roma, Italy
73. Departamento de Física, FCFM, Universidad de Chile, Blanco Encalada 2008, Santiago, Chile
74. Institut de Ciencies de l'Espai (IEEC-CSIC), Campus UAB, Carrer de Can Magrans, s/n Cerdanyola del Vallés, 08193 Barcelona, Spain
75. Centro de Investigaciones Energéticas, Medioambientales y Tecnológicas (CIEMAT), Avenida Complutense 40, 28040 Madrid, Spain
76. Instituto de Astrofísica e Ciências do Espaço, Faculdade de Ciências, Universidade de Lisboa, Tapada da Ajuda, 1349-018 Lisboa, Portugal
77. Universidad Politécnica de Cartagena, Departamento de Electrónica y Tecnología de Computadoras, Plaza del Hospital 1, 30202 Cartagena, Spain
78. Institut de Recherche en Astrophysique et Planétologie (IRAP), Université de Toulouse, CNRS, UPS, CNES, 14 Av. Edouard Belin, 31400 Toulouse, France
79. Kapteyn Astronomical Institute, University of Groningen, PO Box 800, 9700 AV Groningen, The Netherlands
80. Infrared Processing and Analysis Center, California Institute of Technology, Pasadena, CA 91125, USA
81. IFPU, Institute for Fundamental Physics of the Universe, via Beirut 2, 34151 Trieste, Italy
82. AIM, CEA, CNRS, Université Paris-Saclay, Université de Paris, 91191 Gif-sur-Yvette, France
83. Instituto de Astrofísica de Canarias, Calle Vía Láctea s/n, 38204, San Cristóbal de La Laguna, Tenerife, Spain
84. INAF-Istituto di Astrofisica e Planetologia Spaziali, via del Fosso del Cavaliere, 100, 00100 Roma, Italy

[85] Department of Physics and Helsinki Institute of Physics, Gustaf Hällströmin katu 2, 00014 University of Helsinki, Finland
[86] Dipartimento di Fisica e Astronomia "Augusto Righi" – Alma Mater Studiorum Universitá di Bologna, Viale Berti Pichat 6/2, 40127 Bologna, Italy
[87] CEA Saclay, DFR/IRFU, Service d'Astrophysique, Bât. 709, 91191 Gif-sur-Yvette, France
[88] Junia, EPA department, 41 Bd Vauban, 59800 Lille, France
[89] INFN-Bologna, Via Irnerio 46, 40126 Bologna, Italy
[90] Instituto de Física Teórica UAM-CSIC, Campus de Cantoblanco, 28049 Madrid, Spain
[91] CERCA/ISO, Department of Physics, Case Western Reserve University, 10900 *Euclid* Avenue, Cleveland, OH 44106, USA
[92] Laboratoire de Physique de l'École Normale Supérieure, ENS, Université PSL, CNRS, Sorbonne Université, 75005 Paris, France
[93] Observatoire de Paris, Université PSL, Sorbonne Université, LERMA, 75014 Paris, France
[94] Astrophysics Group, Blackett Laboratory, Imperial College London, London SW7 2AZ, UK
[95] SISSA, International School for Advanced Studies, Via Bonomea 265, 34136 Trieste TS, Italy
[96] INFN, Sezione di Trieste, Via Valerio 2, 34127 Trieste, TS, Italy
[97] Dipartimento di Fisica e Scienze della Terra, Universitá degli Studi di Ferrara, Via Giuseppe Saragat 1, 44122 Ferrara, Italy
[98] Istituto Nazionale di Fisica Nucleare, Sezione di Ferrara, Via Giuseppe Saragat 1, 44122 Ferrara, Italy
[99] Institut de Physique Théorique, CEA, CNRS, Université Paris-Saclay 91191 Gif-sur-Yvette Cedex, France
[100] NASA Ames Research Center, Moffett Field, CA 94035, USA
[101] INAF, Istituto di Radioastronomia, Via Piero Gobetti 101, 40129 Bologna, Italy
[102] Institute for Theoretical Particle Physics and Cosmology (TTK), RWTH Aachen University, 52056 Aachen, Germany
[103] Institute for Astronomy, University of Hawaii, 2680 Woodlawn Drive, Honolulu, HI 96822, USA
[104] Department of Physics & Astronomy, University of California Irvine, Irvine CA 92697, USA
[105] University of Lyon, UCB Lyon 1, CNRS/IN2P3, IUF, IP2I Lyon, 4 rue Enrico Fermi, 69622 Villeurbanne, France
[106] Department of Astronomy & Physics and Institute for Computational Astrophysics, Saint Mary's University, 923 Robie Street, Halifax, Nova Scotia, B3H 3C3, Canada
[107] School of Physics, HH Wills Physics Laboratory, University of Bristol, Tyndall Avenue, Bristol, BS8 1TL, UK
[108] University Observatory, Faculty of Physics, Ludwig-Maximilians-Universität, Scheinerstr. 1, 81679 Munich, Germany
[109] Ruhr University Bochum, Faculty of Physics and Astronomy, Astronomical Institute (AIRUB), German Centre for Cosmological Lensing (GCCL), 44780 Bochum, Germany
[110] Department of Physics, Lancaster University, Lancaster, LA1 4YB, UK
[111] Univ. Grenoble Alpes, CNRS, Grenoble INP, LPSC-IN2P3, 53, Avenue des Martyrs, 38000 Grenoble, France
[112] Department of Physics and Astronomy, Vesilinnantie 5, 20014 University of Turku, Finland
[113] Centre de Calcul de l'IN2P3/CNRS, 21 avenue Pierre de Coubertin 69627 Villeurbanne Cedex, France
[114] Dipartimento di Fisica, Sapienza Università di Roma, Piazzale Aldo Moro 2, 00185 Roma, Italy
[115] University of Applied Sciences and Arts of Northwestern Switzerland, School of Engineering, 5210 Windisch, Switzerland
[116] INFN-Sezione di Roma, Piazzale Aldo Moro 2, c/o Dipartimento di Fisica, Edificio G. Marconi, 00185 Roma, Italy
[117] Centro de Astrofísica da Universidade do Porto, Rua das Estrelas, 4150-762 Porto, Portugal
[118] Department of Mathematics and Physics E. De Giorgi, University of Salento, Via per Arnesano, CP-I93, 73100 Lecce, Italy
[119] INFN, Sezione di Lecce, Via per Arnesano, CP-193, 73100 Lecce, Italy
[120] INAF-Sezione di Lecce, c/o Dipartimento Matematica e Fisica, Via per Arnesano, 73100 Lecce, Italy
[121] Institute of Space Science, Str. Atomistilor, nr. 409 Măgurele, Ilfov 077125, Romania
[122] Institute for Computational Science, University of Zurich, Winterthurerstrasse 190, 8057 Zurich, Switzerland
[123] Institut für Theoretische Physik, University of Heidelberg, Philosophenweg 16, 69120 Heidelberg, Germany
[124] Université St Joseph; Faculty of Sciences, Beirut, Lebanon
[125] Leiden Observatory, Leiden University, Niels Bohrweg 2, 2333 CA Leiden, The Netherlands
[126] Department of Astrophysical Sciences, Peyton Hall, Princeton University, Princeton, NJ 08544, USA